\documentclass[aos,preprint]{imsart}

\usepackage[T1]{fontenc}
\usepackage[latin9]{inputenc}
\usepackage[english]{babel} 
\usepackage{algorithm,algpseudocode}
\usepackage{mathtools}
\usepackage{amsmath}
\usepackage{amsthm}
\usepackage{amssymb}
\usepackage{graphicx}
\usepackage{footnote}
\usepackage{fixfoot}
\usepackage{refcount}
\usepackage[numbers]{natbib}
\usepackage{hyperref}
\usepackage{cleveref}
\crefformat{footnote}{#2\footnotemark[#1]#3}

\usepackage{macros}
\usepackage{rf}

\usepackage{hyperref}
\usepackage{tabulary}
\usepackage{booktabs}

\floatstyle{ruled}

  \theoremstyle{plain}

\makeatother

\usepackage{xr}
\begin{document}

\begin{frontmatter}
  \title{A Self-Penalizing Objective Function for Scalable Interaction Detection}
  \runtitle{A Self-Penalizing Objective for Interaction Detection}

  \begin{aug}
    \author{\fnms{Keli} \snm{Liu}\thanksref{t1}\ead[label=e1]{keli.liu25@gmail.com}}
    \and
    \author{\fnms{Feng} \snm{Ruan}\thanksref{t1, m2}\ead[label=e2]{ruanfeng2124@gmail.com}}

    \runauthor{Liu and Ruan}

    \affiliation{
    University of California, Berkeley\thanksmark{m2}}

    \address{
    University of California, Berkeley \\
    Berkeley, California 94720 \\
      \printead{e1} \\
      \phantom{E-mail:\ }\printead*{e2} 
      }

    \thankstext{t1}{The two authors contributed equally to this work (alphabetical-order). }

  \end{aug}
  
\begin{abstract}
~
We tackle the problem of nonparametric variable selection with a focus on discovering
interactions between variables. With $p$ variables there are $O(p^s)$
possible order-$s$ interactions making exhaustive search infeasible.
It is nonetheless possible to identify the variables
involved in interactions with only linear computation
cost, $O(p)$. The trick is to maximize a class of parametrized
nonparametric dependence measures which
we call \emph{metric learning objectives}; the landscape of these nonconvex objective functions is sensitive
to interactions but the objectives themselves do not explicitly model interactions. Three properties
make metric learning objectives highly attractive:
\\
\indent \textbullet~~The stationary points of the objective are automatically sparse (i.e. performs selection)---no explicit $\ell_1$ penalization is needed.
\\
\indent\textbullet~~\emph{All} stationary points of the objective exclude noise variables with high probability.
\\
\indent\textbullet~~Guaranteed recovery of all signal variables without needing to reach the objective's global maxima or special stationary points.

The second and third properties mean that all our theoretical results apply in the practical case where one uses gradient ascent to maximize the metric learning objective. While not all metric learning objectives enjoy good statistical power, we design an objective based on $\ell_1$ kernels that does exhibit favorable power: it recovers (i) main effects with $n \sim \log p$ samples, (ii) hierarchical interactions with $n \sim \log p$ samples and (iii) order-$s$ pure interactions with $n \sim p^{2(s-1)}\log p$ samples.

\end{abstract}  



\end{frontmatter}

\section{Introduction}
\label{chap:intro}

Many real life applications are characterized by strong
variable interactions and only weak main effects. Take for example the ``missing hertiability'' problem in medical
genetics. Genome wide association studies (GWAS) have identified many
single nucleotide polymorphisms (SNPs) that are associated with disease
risk. Surprisingly, the main effects of these SNPs account for only
a small fraction of the inheritable component of disease risk (heritability).
This has led to an intense search for the ``missing heritability''.
\citet{lander2012} postulate that much of this missing heritability
may be hiding in interactions between SNPs. In the case of Crohn's
disease, they estimate that the main effects from currently discovered
SNPs account for 21.5\% of heritability and that interactions account
for as much as 80\% of the unexplained heritability. In the face of
such predictions, it is not surprising that a vast effort has been
mobilized to find gene-gene and gene-environment interactions \citep{hunter2005,cordell2009,liu2013}.
Other applications in which interactions play a critical role include
analyses of consumer behavior \citep{agrawal1994fast}, monitoring
for Alzheimer's progression \citep{li2014}, and failure analyses
of power grids \citep{cai2016cascading}. There is a clear imperative
to develop variable selection methods which account for interactions and do not rely on the existence
of strong marginal dependencies between response and features.

In this paper, we consider a binary response $Y \in \{0,1\}$ and feature vector $X \in \mathbb{R}^p$. For such data, a natural first attempt at formulating the interaction selection problem is through a logistic model:
\begin{align*}
\text{logit }\mathbb{P}(Y=1|X) = \alpha + \sum_{i} \gamma_i X_i + \sum_{j < k} \beta_{jk} \cdot g_{jk}(X_j, X_k).
\end{align*}
An $\ell_1$ penalty (or variant) can then be placed on the coefficients $\beta_{jk}$ to induce selection. To model higher order interactions, one can include richer sets of basis functions, e.g. $g_{jkl}(X_j, X_k, X_l)$ for trivariate interactions. This linear model formulation highlights the two great challenges of the interaction selection problem:
\begin{enumerate}
	\item \emph{Problem of Specification}. Variables can interact in unpredictable ways. An informed choice of basis functions $g$ is difficult. Can we detect interactions without explicit specification of how variables interact?
	\item \emph{Problem of Enumeration}. The number of possible interactions grows exponentially in the number of variables. Fitting the above logistic model is computationally challenging. And clearly, the approach does not scale when trying to detect higher order interactions.
\end{enumerate}
Framed as variable selection within a sparse linear model, interaction detection seems an intractable problem. Our thesis is this: these two great challenges can be (mostly) avoided if we stop representing interactions as parameters in linear models.

\subsection{The Metric Learning Objective}

To get around the problems of specification and enumeration, we ditch the linear model and consider a class of objective functions which we call \emph{metric learning} objectives. The population metric learning objective is defined with respect to independent pairs of observations $(X, Y)$ and $(X', Y')$:
\begin{equation}
F\left(\beta; \mathbb{Q}, f, q\right)= \popexpdiff \left( f\Big( \lVert X - X'\rVert_{q, \beta}^q \Big)\right) \label{eq:obj_f}
\end{equation}
where $\lVert X - X'\rVert_{q, \beta}^q = \sum_{j=1}^p \beta_j \cdot \lvert X_j - X_j'\rvert^q$ is a $\beta$-weighted $\ell_q$-norm and for bivariate functions, $\popexpdiff$ denotes the difference of the between and within class expectations w.r.t. $\Q$:
\begin{equation}
\popexpdiff\left(h(X,X')\right) = \mathbb{E}\left(h(X,X')\vert Y\neq Y'\right) - \mathbb{E}\left(h(X,X')\vert Y= Y'\right).
\end{equation}
The distribution $\Q$ used to construct the objective is usually not equal to $\P$, the true distribution of $(X,Y)$\footnote{In practice, $\mathbb{Q}$ will be a reweighted version of $\mathbb{P}$ such that $\mathbb{Q}(Y=0) = \mathbb{Q}(Y=1)$.}. To obtain the sample metric learning objective, $F_n = F\left(\beta; \Q_n, f, q\right)$, we replace $\Q$ with $\Q_n$ a distribution over the data points $\{(x_i, y_i)\}_{i=1}^n$. One maximizes $F$ or $F_n$ over $\beta \in \mathbb{R}_{+}^p$ to learn a metric that minimizes/maximizes the (feature space) distance between observations in the same/different class(es).

There are several reasons why $F$ (for appropriate choices of $\mathbb{Q}$, $f$, and $q$) is a compelling objective for variable selection:
\begin{itemize}
	\item It is nonparametric. If $X \not \perp Y$ under $\mathbb{P}$, $\beta = 0$ will not be an attractive stationary point of $F$. This holds even when $X$ affects $Y$ through higher order interactions or pure interactions where each individual feature $X_j \perp Y$ ($1\leq j\leq p$).
	\item It is computationally cheap. We maximize $F$ with (stochastic) gradient ascent. We show that even accounting for the number of iterations needed for convergence, the total computation cost is still linear in $p$. Finding higher order pure interactions usually involves a combinatorial search through subsets of variables but not if we use $F$.
	\item It is nonconvex but the success of our variable selection algorithm will not require finding global optima of $F$. Our statistical guarantees assume only that we reach a stationary point of $F$.
	
	\item It performs variable selection without need for explicit penalization (e.g. $\ell_1$) of $\beta$. The stationary points of the finite sample metric learning objective, $F_n$, are automatically sparse w.h.p. whenever $X \not \perp Y$.
\end{itemize}
Our first contribution will be to elucidate the $\mathbb{Q}$, $f$, and $q$ which endows the metric learning objective with these nice properties.

Our second contribution will be to show how to actually use $F$ (and its finite sample version $F_n$) to build a provably correct variable selection algorithm. Surprisingly, the support of the maximizer, $\beta^* = \argmax_{\beta} F(\beta)$, does not coincide with the set of signal variables. In fact, the coefficients $\beta_j$ corresponding to features with weak signals may be 0 at \emph{all} stationary points of $F$. The correct way of using $F$ is via a sequential algorithm that maximizes a series of reweighted objectives $F(\beta; \Q^{(k)})$. This algorithm iterates between two steps:
\begin{enumerate}
\item Find any stationary point of $F(\beta; \Q^{(k)})$ and add its support to the set of selected variables, $\hat S$.
\item Construct $\Q^{(k+1)}$ so that $Y \perp X_{\hat S}$ under $\Q^{(k+1)}$ but $\Q^{(k+1)}$ retains the relationship between $Y$ and signal variables we have not yet discovered.
\end{enumerate} 
The culmination is the metric screening algorithm, Algorithm \ref{alg:pop}---a fully nonparametric variable selection algorithm, capable of finding even pure interactions with a computation cost that is linear in the number of variables. Algorithms \ref{alg:metric_screen_low_dim} and \ref{alg:metric_screen_high_dim} are finite sample variants of Algorithm \ref{alg:pop} appropriate for use with low and high dimensional data respectively. Both come equipped with provable false positive and recovery guarantees.

\begin{algorithm}[!tph]
\begin{algorithmic}[1]
\Ensure{$\hat S=\emptyset$ and $\Q$ s.t. $\frac{d\Q}{d\P} = w(x,y)$ where $w(x, y) = \P(Y=1-y)$}
\While{ $F(\frac{1}{p}\mathbf{1}_p; \Q) \neq 0$}
\State{Find any stationary point $\beta$ of $F(\beta; \Q)$ and set $\hat S=\hat S \cup \supp(\beta)$.}
\State{Update $\Q$ s.t. $\frac{d\Q}{d\P} = w(x,y)$ where $w(x,y) = \P(Y=1-y|X_{\hat S} = x_{\hat S})$.} 
\EndWhile
\end{algorithmic}
\caption{Population Algorithm}
\label{alg:pop}
\end{algorithm}


\subsection{Relation to Prior Work}

To bypass the computational difficulties of detecting interactions, the literature has relied heavily on the \emph{hierarchy} assumption: a higher order interaction exists only if a subset of the interacting variables exhibits a lower order interaction. For example, two variables can interact only if at least one of the variables is marginally dependent with the response\footnote{This corresponds to weak hierarchy. Variations exist, e.g. strong hierarchy, where both variables must have a main effect.}. To discover hierarchical interactions, one can proceed in a stagewise manner: first find variables that are marginally dependent with the response, then look for two way interactions but limiting our search to interactions that involve one of the discovered variables, and so on. Such a stagewise approach can be explicit, often the case when linear models are involved~\cite{Kooperberg08, WuChHaSoLa09, WuDeRiTrRo10, HaoZh17, HaoFeZh18}. Or it can be implicit, e.g. in tree based variable selection methods \citep{loh2002regression,strobl2008conditional,archer2008empirical,genuer2010variable,yu2018iterative} where trees are trained via a greedy algorithm like CART---to capture a two way interaction, the tree must first split on one of the interacting variables, which, excluding sheer luck, presumes that one of the variables has a main effect. The great benefit of the stagewise approach is that it allows for finding hierarchical interactions with a computational cost linear in $p$.

The metric screening algorithm allows us to match the linear in $p$ computation cost \emph{without} having to assume that interactions are hierarchical. It is almost magical that one can detect a pure two way interaction (both variables are independent of the response) without having to consider all $\binom{p}{2}$ possibilities. The magic is hiding in the metric learning objective, $F$, special cases of which has appeared in prior work. When $f(x) = \sqrt{x}$ and $q=2$, $F$ becomes distance covariance \citep{szekely2009brownian}; when $f(x) = -e^{-x}$ and $q=2$, $F$ reduces to the maximal mean discrepancy (MMD) induced by a Gaussian kernel\footnote{The equivalence holds for balanced class sizes.}. Both distance covariance and MMD are well known nonparametric measures of dependence between $X$ and $Y$. So maximizing $F$ can often be interpreted as maximizing the dependence between the response and features, an idea also employed by \citet{song2012feature}.

While it may seem intuitive to select variables by maximizing some measure of dependence between $X$ and $Y$, the results of the maximization are often quite counter-intuitive! As we discussed above, simply maximizing $F$ leads to an incorrect variable selection algorithm. This is a crucial point that has been overlooked by the literature where dependence maximization methods are often justified only through heuristics and perhaps a warning that ``a local optimum does not necessarily provide good features'' (the global optimum doesn't either!). These shortcomings stem from the absence of an effective technique for analyzing objectives like $F$. Using Fourier analytic tools, this paper offers the following:
\begin{itemize}
\item \emph{A systematic look at the optimization landscape of nonparametric dependence maximization}. We characterize the stationary points of $F$ and this understanding allows us to build an algorithm which does not require one to find global optima or special stationary points.
\item \emph{Design considerations for objectives based on nonparametric dependence measures}. When $f(x) = -e^{-x}$, the choice of either $q=1$ or $q=2$ leads to $F$ being a valid nonparametric dependence measure. However, the sample gradients of $F_n$ when $q=1$ has much better statistical properties. Not all nonparametric dependence measures are equally good for variable selection.
\end{itemize}
Our great hope is that by the end of our analysis, the metric learning objective (and its various cousins based on other nonparametric dependence measures) will have become a bit less black-box to the reader. And variable selection via dependence maximization, a bit less in common with the Wild West.

\subsection{Problem Setup}
\label{subsec:def_noise}

We consider a binary response $Y \in \{0,1\}$ and feature vector $X \in \mathbb{R}^p$.
For any joint distribution $\P$ on $(X,Y)$, there exists a \emph{unique and minimal} subset $S\subseteq\left\{ 1,\ldots,p\right\} $, such that (i) $Y|X\sim Y|X_{S}$ and (ii) $X_S \perp X_{S^c}$ (see Proposition \ref{prop:uniq_noise} in Appendix \ref{sec:basics-on-models})---this is not an assumption! We call $S$ the set of signal variables and $S^c$ the set of noise variables\footnote{The symbols $S$ and $S^c$ are reserved for this purpose.}. Our goal is to recover a subset $\hat S \subseteq S$ such that $Y|X \sim Y|X_{\hat S}$ and do so with a provably correct algorithm whose computation cost can be made linear in $p$, the number of features. We make no further distributional assumptions about $(X,Y)$. We list and address the most common objections to our setup:

\begin{itemize}
	\item \emph{Are you assuming that signal and noise variables are independent?} Exact independence between signal and noise variables is admittedly a strong assumption: in variable selection for linear models, one only constrains the correlation between noise and signal variables. The stronger assumption is the price we pay for working in a nonparametric setting. Without such independence assumptions, we can still make statements about power i.e. our algorithm returns a set $\hat S$ such that $Y|X \sim Y|X_{\hat S}$ but we cannot say anything about false positive control.
	\item \emph{Why can't you guarantee $\hat S =S$? Why can't you find the smallest $\hat S \subseteq S$ such that $Y|X_{\hat S} \sim Y|X$?} Due to dependence between variables in $S$, there may not exist a unique way to describe the relationship between $Y$ and $X$, e.g., if $S=\{1, 2\}$ and $X_{1}=X_{2}$. Many subsets $A\subseteq S$ may exist all satisfying $Y|X \sim Y|X_A$. Our algorithm will return one such subset but we can't guarantee which\footnote{Nor can we guarantee that the set $\hat S$ we return is somehow the most parsimonious. Suppose $\mathbb{E}[Y|X]=\pi(X_{1})$ and $X_2$ is a noisy version of $X_1$. Ideally, we'd return $\hat S =\{1\}$. But since $Y$ is marginally dependent with $X_2$, we may add variable $2$ to $\hat S$ even though it provides no additional information given $X_1$. This situation is unlikely to happen because the algorithm we propose tends to select the strongest variables first (see Section \ref{subsec:maineff_ex}).}. Only with further identifiability assumptions, e.g. each signal variable carries non-redudant information, can we guarantee $\hat S = S$ (see Proposition \ref{cor:dependent_main} in Appendix~\ref{sec:proofs_all_recovery}).
\end{itemize}

\subsection{Outline of Paper}

The paper is organized as follows:
\begin{itemize}
\item Section \ref{chap:f_restrict} derives properties of the population metric learning objective and proves correctness of the population metric screening algorithm.
\item Section \ref{chap:finite_sample_alg} introduces two finite sample metric screening algorithms, one suited to low dimensional data (Algorithm \ref{alg:metric_screen_low_dim}) and the other high dimensional data (Algorithm \ref{alg:metric_screen_high_dim}). 
\item Section \ref{chap:exclude_noise} shows that all stationary points of the sample objective $F\left(\beta; \Q_n\right)$ exclude noise variables with high probability. False positive control is achieved via self-penalization in Algorithm \ref{alg:metric_screen_low_dim} and via explicit $\ell_1$ penalization in Algorithm~\ref{alg:metric_screen_high_dim}. 
\item Section \ref{chap:recovery} provides recovery results. In low dimensions, we prove that the selected variables, $\hat S$, satisfy $Y|X_{\hat S} \sim Y|X_S$ for all signals. In high dimensions, we compute the sample size requirements for recovering specific signal types: main effects ($n \sim \log p$), hierarchical interactions ($n \sim \log p$), and pure interactions ($n\sim p^{2\left(s-1\right)}\log p$). Our recovery results do not require finding global maxima.
\item Section \ref{chap:simulatons} provides experimental validation of theoretical
results.
\end{itemize}

\paragraph{Notation}


\begin{table}[htbp]
\begin{center}
\begin{tabular}{r c p{10cm} }
\toprule
$\R$      & $\triangleq$ &the set of real numbers\\
$\R_+$ & $\triangleq$ & the set of nonnegative reals \\
$\N$ & $\triangleq$ & the set of positive integers\\
$\N_0$ & $\triangleq$ & the set of nonnegative integers \\
$[d]$ & $\triangleq$ &the set $\{1, 2, \ldots, d\}$ \\
$v_S$ & $\triangleq$ & restriction of a vector $v$ to the index set $S$\\
$\supp(v)$ & $\triangleq$ & the support set $\{j: v_j \neq 0\}$ for a vector $v$ \\
$\norm{v}_p$ & $\triangleq$ & the $\ell_p$ norm of the vector $v$: $\norm{v}_p = (\sum_i |v_i|^p)^{1/p}$.\\  
$\norm{v}_{p, \beta}$ & $\triangleq$ & the weighted $\ell_p$ norm of the vector $v$: 
			$\norm{v}_{p, \beta} = (\sum_i \beta_i |v_i|^p)^{1/p}$\\  
$\mathcal{C}^\infty(\R_+)$  & $\triangleq$ & the set of functions $f$ infinitely differentiable 
				 on $\R_{++} = (0, \infty)$ whose derivatives are right continuous at $0$ \\
$f^{(l)}(x)$ & $\triangleq$ & the $l$-th derivative of a function $f$ at $x$ \\
$f^{(l)}(0)$ & $\triangleq$ & $\lim_{x\to 0^+} f^{(l)}(0)$ for any $f \in \mathcal{C}^\infty(\R_+)$ \\
$f(+\infty)$ & $\triangleq$ & $\lim_{x \to +\infty} f(x)$ when the limit exists \\
$\Q_1 \ll \Q_2$ && if the measure $\Q_1$ is absolutely continuous w.r.t the measure $\Q_2$. \\
$a_n = O(b_n)$ && if $\limsup_{n \to \infty} \frac{|a_n|}{b_n} < \infty$. Here $b_n > 0$. \\
$a_n = \Omega(b_n)$ && if $\liminf_{n \to \infty} \frac{a_n}{b_n} > 0$. Here $b_n > 0$.  \\
$a_n \sim b_n$ && if $a_n = O(b_n)$ and $a_n = \Omega(b_n)$. Here $a_n, b_n > 0$. \\
\bottomrule
\end{tabular}
\end{center}
\end{table}


The notation $\P$ is reserved for the true distribution of the data $(X, Y)$. For a generic probability distribution $\Q$, we use $\E_B$ and $\E_W$ to denote the expectation 
of the between group and of the within group pairs under $\Q$. More precisely, 
for any function $h(x, x', y, y')$, we define 
\begin{equation*}
\begin{split}
\E_B[h(X, Y, X', Y')] &= \E[h(X, Y, X', Y') \mid Y \neq Y'], \\
\E_W[h(X, Y, X', Y')] &= \E[h(X, Y, X', Y') \mid Y = Y'].
\end{split}
\end{equation*}
where $(X, X')$, $(Y, Y')$ are independent pairs sampled from $\Q$.
We use $\E_{B-W}$ as a notational shorthand for $\E_B - \E_W$.

For a generic probability distribution 
$\Q \ll \P$, we use $\Q_n$ to denote the weighted empirical measure with weights 
$w(x, y) \propto \frac{d\Q}{d\P}(x, y)$ where $\frac{d\Q}{d\P}(x, y)$ is the Radon-Nikodym derivative 
of $\Q$ w.r.t $\P$. We use $\hat{\E}_n$ to denote the expectation under $\Q_n$: 
for any function $h(x, y)$, this becomes
\begin{equation*}
\hat{\E}_n(h(X, Y)) = \frac{\sum_{i=1}^n h(X_i, Y_i) w(X_i, Y_i)}{\sum_{i=1}^n w(X_i, Y_i)},
\end{equation*}
where $(X_i, Y_i)$ are i.i.d data drawn from $\P$. Analogous to $\E_B, \E_W, \E_{B-W}$, 
we can define $\empexpB, \empexpW, \empexpdiff$. For any function $h(x, x', y, y')$,
\begin{equation*}
\begin{split}
\empexpB[h(X, X', Y, Y')] &=
	 \frac{ \sum_{1\le i, i' \le n} w(X_i, Y_i) w(X_{i^\prime}, Y_{i^\prime})
	 	h(X_i, X_{i^\prime}, Y_i, Y_{i^\prime}) \mathbf{1}_{Y_i \neq Y_{i^\prime}}} 
	{\sum_{1\le i, i' \le n} w(X_i, Y_i) w(X_{i^\prime}, Y_{i^\prime})\mathbf{1}_{Y_i \neq Y_{i^\prime}}} \\
\empexpW[h(X, X', Y, Y')] &=
	 \frac{ \sum_{1\le i, i' \le n} w(X_i, Y_i) w(X_{i^\prime}, Y_{i^\prime})
	 	h(X_i, X_{i^\prime}, Y_i, Y_{i^\prime}) \mathbf{1}_{Y_i = Y_{i^\prime}}} 
	{\sum_{1\le i, i' \le n} w(X_i, Y_i) w(X_{i^\prime}, Y_{i^\prime})\mathbf{1}_{Y_i = Y_{i^\prime}}}
\end{split},
\end{equation*}
where $w(x, y) \propto \frac{d\Q}{d\P}(x, y)$. Finally, we can define the population and empirical metric learning objectives w.r.t. $\Q$ by
\begin{equation*}
\begin{split}
F(\beta; \Q, f, q) &= \E_{B-W}[f(\norm{X- X'}_{q, \beta}^q)] \\
F(\beta; \Q_n, f, q) &= \hat{\E}_{n, B-W}[f(\norm{X- X'}_{q, \beta}^q)].
\end{split}
\end{equation*}

We will also use the following basic definitions and conventions: 
\begin{enumerate}[(i)]
\item Consider the following (possibly non-convex) optimization problem: 
	\begin{equation*}
		\maximize_{\beta \in \mathcal{C}} g(\beta)
	\end{equation*}
	where $g: \R^p \to \R$ is smooth and $\mathcal{C}$ is convex. 
	A point $\beta \in \mathcal{C}$ is called a~\emph{stationary point} w.r.t the optimization 
	problem if $\langle \nabla g(\beta), \beta' - \beta \rangle \le 0$ for all $\beta^\prime \in \mathcal{C}$~\cite{Bertsekas97}.
	The \emph{projected gradient ascent} algorithm with initialization $\beta^{(0)}$ and stepsize $\alpha$ proceeds 
	as follows: 
	\begin{equation*}
		\beta^{(k+1)} = \Pi_{\mathcal{C}} (\beta^{(k)} + \alpha \nabla g(\beta^{(k)})).
	\end{equation*}
	Here $\Pi_{\mathcal{C}}$ is the $\ell_2$ projection operator: 
	$\Pi_{\mathcal{C}}(\beta) = \argmin_{\beta' \in \mathcal{C}} \ltwo{\beta' - \beta}$.
\item A random variable $X$ is non-degenerate if its distribution is not a point mass. 
	To avoid unnecessary technicality, we assume throughout the paper that all random variables are non-degenerate. 
\end{enumerate}

\section{Designing the Population Objective}

\label{chap:f_restrict}

In this section we consider the population metric screening algorithm (Algorithm \ref{alg:pop}) and show that it selects the correct variables. The population algorithm maximizes a sequence of objectives,
\[
F\Big(\beta; \Q^{(k)}\Big)=\popexpdiff\left[f\big(\norm{X-X'}_{q, \beta}^q\big)\right],
	~~k=1,2,\ldots,
\]
and selects $\hat S = \cup_k  \supp(\hat{\beta}^{(k)})$ where $\hat{\beta}^{(k)}$ is the result of running gradient ascent on the $k$-th objective. To endow the selected variables, $\hat S$, with the appropriate properties, we must choose the right $f$ and use the right sequence of distributions $\Q^{(k)}$. We establish that:
\begin{enumerate}
\item The maximizer of $F\big(\beta; \Q\big)$ will have support containing (i) at least one non-null variable and (ii) no null variables 
(under $\Q$), if and only if the derivative $f'$ is a strictly completely monotone function.
\item A finite sequence $\Q^{(k)}$ can be constructed such that every non-null variable (w.r.t. the true data distribution $\P$) appears in the support of the maximizer of $F(\beta ; \Q^{(k)})$ for some $k$.
\end{enumerate}
Putting these two facts together gives a population algorithm that selects all non-null variables and no null variable (see Section \ref{subsec:def_noise} for definitions of null and non-null). Sections~\ref{chap:exclude_noise} and \ref{chap:recovery} will build on this population level analysis to characterize the false positive control and power properties of the \emph{finite sample} metric screening algorithm.

\subsection{An Axiomatic Derivation of $f$}

\label{subsec:f_key_prop}

We will use two common sense axioms to guide our choice of $f$. These two axioms force $f'$ to be a completely monotone function and turns $F(\beta; \Q)$ into a nonparametric measure of dependence between $X$ and $Y$ (under $\Q$).

\begin{axiom}
\label{axiom:XOR}
The objective $F(\beta; \Q) > 0$ whenever $X_{\supp(\beta)}$ and $Y$ exhibit an XOR-type dependence under $\Q$.
\end{axiom}

\begin{axiom} 
\label{axiom:translation-invariant}
Let $F_{c}(\beta; \Q) = \E_{B-W}[f(c + \norm{X-X'}_{q, \beta}^q)]$ where $c \ge 0$. A function $f$ is 
called translation invariant if for all $c \ge 0$,
\begin{equation*}
	\text{$F_0(\beta; \Q) >0$ implies $F_{c}(\beta; \Q) >0$}.
\end{equation*}
\end{axiom}


\begin{remark}
$(X_1,\ldots,X_s)$ and $Y$ exhibit an order-$s$ XOR signal if $Y=1_{X_{1}X_{2}\ldots X_{s}>0}$ and $X_{j}$ are
i.i.d random signs with $\Q\left(X_{j}=\pm\frac{1}{2} \right)=\frac{1}{2}$.  An order-$s$ XOR signal is a type of pure interaction, which as we know, is incredibly difficult to detect. Since the focus of this paper is on detecting arbitrary interactions, we should obviously force $F(\beta; \Q)$ to be sensitive towards XOR-type signals. Axiom 1 does precisely this. One could replace Axiom 1 with a stronger version: 
$F(\beta, \Q) > 0$ whenever $X_{\supp(\beta)}$ and $Y$ are dependent (under $\Q$). Surprisingly, the stronger axiom is unnecessary! 
The set of XOR signals
is sufficiently rich that if we choose $f$ with the ability to detect
all XOR signals, such an $f$ automatically detects \emph{any} departure from independence between $X$ and $Y$.
\end{remark}

\begin{remark}
Axiom 2 arises from the need to detect signals in the presence
of noise variables. Suppose the variable set $\left\{ 1,\ldots,p\right\} $
can be partitioned into subsets $S$ (signal) and $S^{c}$ (noise) such that $X_{S}\perp X_{S^{c}}$
and $Y\perp X_{S^{c}}|X_{S}$ (under $\Q$). Then we have the identity 
\begin{equation}
F(\beta; \Q)
= \E\left[\popexpdiff\left[f\left(c(X_{S^{c}},X_{S^{c}}^\prime)+\lVert X_{S}-X_{S}'\rVert_{q}^{q}\right) \mid X_{S^{c}},X_{S^{c}}^\prime \right]\right]
\label{eq:noise_intercept}
\end{equation}
where $c(X_{S^{c}},X_{S^{c}}^\prime)=\lVert X_{S^{c}}-X_{S^{c}}'\rVert_{q}^{q}$. We see that
independent noise variables essentially contribute an ``intercept''
term to the metric learning objective. Our ability to detect signal should not be reliant
on a particular choice of intercept for the function $f$---this leads us to choose translation invariant $f$.
\end{remark}
 
We now explore the implications of our two common sense axioms. We derive an integral representation for $F(\beta; \Q)$ when $X$ and $Y$ exhibit an XOR-type signal. The restrictions on $f$ implied by Axiom 1 can then be read off directly from the integral representation
(see Lemma~\ref{lem:general_xor} and its proof in Appendix~\ref{sec:proof_f_restrict}). 
\begin{lemma}
\label{lem:xor_uniform}For any function $f: \R_+ \mapsto \mathbb{R}$
that is $\mathcal{C}^{s}$ continuously differentiable in $\R_+$, we have for any $c\in\mathbb{R}_{+}$, 
\begin{align}
\label{eqn:xor_uniform}
F_{c}(\beta) 
 =\left(-1\right)^{s-1}\int_{0}^{1}\int_{0}^{1}\ldots\int_{0}^{1}f^{(s)}\bigg(c+\sum_{k=1}^{s} \beta_k t_{k}\bigg)dt_{1}dt_{2}\ldots dt_{s}
\end{align}
when $(X_1,\ldots,X_s)$ and $Y$ exhibit an order-$s$ XOR signal.
\end{lemma}
Using equation~\eqref{eqn:xor_uniform}, we see that in order for $F(\beta; \Q) > 0$ for any 
XOR signal and any $c > 0$, we need $(-1)^{s-1}f^{\left(s\right)}\left(x\right) > 0$
for all $s \ge 1$ i.e. that the derivative $f'$ is a strictly completely monotone function.

\begin{definition}
A function $f: \R_+ \mapsto \R$ is called completely monotone if 
(i) $f \in \mathcal{C}^\infty (\R_+)$ and $(ii)$ for all $k\in \N_0$ and $x \in \R_+$,
\begin{equation}
\label{eqn:def-cmfunction}
	(-1)^{k} f^{(k)}(x) \ge 0. 
\end{equation}
A completely monotone function $f: \R_+ \mapsto \R$ is called strictly completely monotone if the 
inequality~\eqref{eqn:def-cmfunction} is strict for all $k \in \N_0$ and $x \in \R_+$.
\end{definition}

As an concrete example, consider
$f(x)=-(p-x)^{s}$ (this choice is not translation
invariant). Then $(-1)^{k-1}f^{(k)}(x) > 0$ for all $k$ up to $s$ and $x \in [0, p)$.
Hence, this choice of $f$ allows us to detect XOR signals of order $s$ or less. If we wish to
detect higher order interactions, we must make $f$ more nonlinear
i.e. go beyond an $s$ degree polynomial. Two popular choices are
$f(x) = -e^{-x}$ and $q=2$ which induces a
Gaussian kernel and leads to the MMD statistic \citep{song2012feature}
and $f(x) = \sqrt{x}$ and $q=2$ which leads to a distance
covariance type measure \citep{GaborMa2005}. Both these choices are
translation invariant, satisfy equation \eqref{eqn:def-cmfunction} for all their
derivatives, and lead to nonparametric measures of dependence. As the following theorem shows, 
this last property is no coincidence (the proof of Theorem~\ref{thm:f_condition} is in 
Appendix~\ref{sec:proof_f_restrict}).

\begin{theorem}
\label{thm:f_condition}
Suppose $f \in \mathcal{C}^{\infty}(\R_+)$ satisfies Axioms \ref{axiom:XOR}-\ref{axiom:translation-invariant}. Then, 
\begin{itemize}
\item $f'$ is strictly completely monotone on $\R_+$, i.e.,  for all $k \in \N$, 
	\begin{equation*}
		(-1)^{k-1} f^{(k)}(x) > 0~~\text{for $x \in \R_+$}.
	\end{equation*}
\item For $q \in \{1, 2\}$, $F(\beta) \ge 0$ with equality if and only if $X_{\supp(\beta)} \perp Y$.
\end{itemize}
\end{theorem}

Based on Theorem \ref{thm:f_condition}, we can interpret the population algorithm as maximizing a sequence of nonparametric 
dependence measures $F(\beta; \Q^{(k)})$. The solution to each subproblem in the sequence has a nice property: we are 
guaranteed to exclude all noise variables and to recover at least one signal variable (see Proposiiton~\ref{prop:obj_nice_property};
proof in Appendix~\ref{sec:basics-on-algorithm}).



\begin{proposition}
\label{prop:obj_nice_property}
Let $f \in \mathcal{C}^\infty(\R_+)$ be such that $f^\prime$ is strictly completely monotone. 
Let $q \in \{1, 2\}$. If $\beta$ is a local maxima of $F(\beta)$ with $F(\beta) > 0$, then we must have
\begin{itemize}
	\item $X_{\supp(\beta)}$ and $Y$ are dependent.
	\item $Y | X_{\supp(\beta)} \not\sim Y|X_A$ for any $A \subsetneq \supp(\beta)$
	such that $X_A \perp X_{\supp(\beta) \backslash A}$.
\end{itemize}
\end{proposition}
 
 Unfortunately, Proposition~\ref{prop:obj_nice_property} does not guarantee that the solution to any one subproblem will include all signal variables (in fact, a counterexample exists regardless of the choice of $f$). In Section~\ref{subsec:maineff_ex} we show what goes wrong when selecting variables by naively maximizing a nonparametric dependence measure and in Section~\ref{subsec:rebalance}, we show how to fix it.

\subsection{The Masking Phenomenon}

\label{subsec:maineff_ex}

In this section, we show that stationary points of $F(\beta)$ are not guaranteed to contain all signal variables i.e. $Y|X \not\sim Y|X_{\supp(\beta)}$. One might be tempted to ask the following questions:
\begin{itemize}
\item Does the global maximizer of $F(\beta)$ contain all signal variables?
\item Are all (or most) signal variables present in at least one of the stationary points of $F(\beta)$?
\end{itemize}
Unfortunately, the answers to both questions are no. Even if we could
visit all stationary points of $F(\beta)$, we would \emph{still}
miss out on important variables. The culprit behind the selection inconsistency of $F(\beta)$ is
a phenomenon we call \emph{masking}: variables compete against one
another and stronger variables prevent weaker variables from being
selected.

We demonstrate the masking phenomenon with a concrete example
where all the stationary points of $F(\beta)$ can be computed
explicitly.


\begin{example*}
Consider the following additive main effect model:
\begin{equation*}
\begin{split}
	&\Q(Y = 1) = \Q (Y = 0) = \half, ~~X_{1}\perp X_{2}\perp\ldots\perp X_{s}|Y, \\
	&\Q(X_j = \pm \half \mid Y = 1) = \half (1 \pm \delta_j)~~\text{for $j \in [s]$},  \\
	&\Q(X_j = \pm \half \mid Y = 0) = \half (1 \mp \delta_j)~~\text{for $j \in [s]$}. \\
\end{split}
\end{equation*}
Under this model, each binary feature $X_j$ has signal size $\delta_j$. 
For interpretability of results below, we reparametrize the signal size as 
$\rho_j = (1+\delta_j^2)/(1-\delta_j^2)$. Assume $\rho_1 > \rho_2 \ldots > \rho_s > 0$. 
Fix $f(x)=-e^{-x}$. Roughly speaking, the stationary points of $F(\beta; \Q)$ 
for $q > 0$ exhibits the following properties (Proposition \ref{prop:ksparse_loc_minima} in 
Appendix \ref{sec:landscape} makes rigorous the qualitative description below):
\begin{itemize}
\item[--] $\beta=0$ is never a stationary point when main effects are present.
\item[--] If there is a dominant effect, i.e. $\rho_{1}\gg\rho_{j}$ for $j\neq1$,
$F(\beta; \Q)$ has a single stationary point at 
$\left(\infty,0,\ldots,0\right)$\footnote{$(\infty, 0, \ldots, 0)$ is called a stationary point of $F(\beta; \Q)$ 
if $(r, 0, \ldots, 0)$ is a stationary point of $F(\beta; \Q)$ with respect to the 
box constraint $\{\beta \in \R_+^p: 0 \le \beta_i \le r\}$ for all large enough $r$.}.
The dominant effect masks all weaker effects!
Even if $\rho_{1}$ is not dominant, $\left(\infty,0,\ldots,0\right)$ remains the only
$1$-sparse stationary point but there may be other $k$-sparse stationary
points for $k>1$.
\item[--] No 2-sparse stationary points are feasible. One of the effects in
a 2-sparse solution will always be stronger (by assumption) and will
mask the weaker effect.
\item[--] For $k>2$, a $k$-sparse stationary point $\beta$ with $\supp(\beta) = \left\{ j_{1},\ldots,j_{k}\right\} $
exists only if the effect sizes $\rho_{j_{1}},\ldots,\rho_{j_{k}}$ are
roughly equal and if effects $\rho_{l}$ for $l$ outside the support are not too strong.
\end{itemize}
The last point implies the stationary points of $F(\beta; \Q)$ tend to be highly sparse ($\left|\beta\right|_{0}$ is small) 
unless our variables have relatively uniform effect sizes, which is unlikely in practice. 
\end{example*}

In our simple model with binary predictors and independent main effects,
we are able to explicitly characterize all the stationary points of
$F(\beta; \Q)$. The chief conclusion is that weaker predictors
are unlikely to be selected by simply minimizing $F(\beta; \Q)$.
In fact, for any choice of strictly completely monotone $f'$, one can construct an example where important variables do not appear in the support of any of $F(\beta; \Q)$'s stationary points (see Proposition \ref{prop:f_inconsistency}; proof in Appendix~\ref{sec:proof-prop-f_inconsistency}). The counterexample should put an end to
an attractively simple (and commonly used) heuristic for variable selection---maximize some nonparametric measure of dependence between $Y$ and $X$. This heuristic, by itself, is wrong. The next subsection takes the heuristic of dependence maximization and shows how to use it within a provably correct variable selection algorithm.

\begin{proposition}
\label{prop:f_inconsistency}
We have the below inconsistency result. Consider
\begin{equation*}
	F(\beta; \Q) = \popexpdiff[f\big(\norm{X-X'}_{q, \beta}^q\big)],
\end{equation*}
where $f \in \mathcal{C}^\infty(\R_+)$ is such that $f^\prime$ is strictly completely monotone, $f^\prime(\infty) = 0$ 
and $q > 0$. Let $\mathcal{D}_r = \left\{\beta \in \R^p: 0\le \beta_i \le r\right\}$ denote the box-constraint set.
\begin{enumerate}
\item There exists a distribution $\Q$ such that, for all large enough $r$, any global maximizer $\beta^*$ of 
	the objective $F(\beta; \Q)$ over the constraint set $\mathcal{D}_r$ can't satisfy $Y|X \sim Y|X_{\supp(\beta^*)}$. 
\item Assume $f^\prime$ is not slowly varying at 
	$+\infty$\footnote{A function $f: \R_+ \to \R$ is \emph{slowly varying} at $+\infty$ if 
	$\lim_{x \to +\infty} \frac{f(ax)}{f(x)} = 1$ for all $a > 0$.}. Then there exists a 
	distribution $\Q$ such that any stationary point of the objective $F(\beta; \Q) $ w.r.t.
	the constraint $\mathcal{D}_r$ can't have $Y|X \sim Y|X_{\supp(\beta^*)}$ for all large enough $r$. 
\end{enumerate}
\end{proposition}

\subsection{Rebalancing}

\label{subsec:rebalance}

Due to the masking phenomenon, \emph{even if} we could find the global maxima of $F(\beta; \Q)$, we still wouldn't be able to identify all important variables. Fortunately, we can reconstruct the full set of important variables by finding stationary points for a sequence of reweighted objectives $F(\beta; \Q^{(k)})$. The intuition for maximizing a sequence of objectives is simple: to recover variables whose (relatively) weaker effects are being masked, we will need to remove the effect of the stronger variables. 

We do so by reweighting the distribution $\P$ into a new distribution $\Q$ with the following properties:
\begin{itemize}
\item If $\hat{S}$ is the set of variables selected thus far, $X_{\hat{S}} \perp Y$
under $\Q$. Selected variables no longer affect $Y$.
\item The conditional distribution of $X_{\hat{S}^{c}}|X_{\hat{S}},Y$ is
the same under $\P$ as under $\Q$. Signal in $\hat{S}^{c}$
but \emph{not} in $\hat{S}$ is preserved.
\item Noise variables under $\P$ remain noise under $\Q$. Since noise and signal are independent in our setup, this follows from the previous property. 
\end{itemize}
The operation $\P \rightarrow \Q$ which achieves these properties is called rebalancing and is defined by:
\begin{equation*}
\frac{d\Q}{d\P}\left(x,y\right)\propto w(x,y),~~\text{where}~~
w(x,y) = \P\left(Y=1-y|X_{\hat{S}}=x_{\hat{S}}\right).
\end{equation*}
Propositions \ref{prop:balance} gathers the basic properties of rebalancing. 
\begin{proposition}
\label{prop:balance}
The following properties hold for rebalancing: 
\begin{enumerate}[(a)]
\item The variables $X_{\hat{S}}$ and $Y$ are independent under $\Q$. $\Q(Y = 1) = \half$.
\item The conditional distribution $X_{\hat{S}^{c}} \mid X_{\hat{S}},Y$ is the same 
	under $\Q$ as under $\P$. Hence, if $X_{\hat{S}} \perp X_{\hat{S}^c} \mid Y$ under $\P$, 
	then $X_{\hat{S}} \perp X_{\hat{S}^c} \mid Y$ under $\Q$. 
\item $Y | X = Y | X_S$ and $X_S \perp X_{S^c}$ under $\Q$. 
\end{enumerate}
\end{proposition}

The rebalanced distribution $\Q$ induces an objective $F(\beta; \Q)$ with the following key property: if $\hat S$ already contains all signal variables (w.r.t $\P$), then $F(\beta; \Q) = 0$ for any $\beta$, otherwise, any non-zero stationary point of $F(\beta; \Q)$ must be a non-empty subset of the set of \emph{undiscovered} true variables (Proposition \ref{prop:reweight_obj}). Since the stationary points of the rebalanced objective contain undiscovered variables, iterative rebalancing with respect to the set of selected variables allows one to eventually find all true variables. This logic forms the foundation of the population algorithm (Algorithm \ref{alg:pop}). To know when to terminate the algorithm, note that $F(\beta; \Q) > 0$ for all $\beta > 0$\footnote{This means that $\beta_i > 0$ for all $i \in [p]$.} whenever $X \not \perp Y$ (Theorem \ref{thm:f_condition}); hence $F(\frac{1}{p}\mathbf{1}_p; \Q) = 0$ implies $X \perp Y$ under $\Q$ which implies $Y|X \sim Y|X_{\hat S}$.

\begin{proposition}
\label{prop:reweight_obj}
Let $\hat S \subseteq S$ 
and $w = \P\left(Y=1-y|X_{\hat{S}}=x_{\hat{S}}\right)$. Let $\P^w$ be the unique probability distribution such 
that $dP^w(x, y) \propto dP(x, y) \cdot w(x, y)$.
\begin{itemize}
	\item If $Y|X_S \not \sim Y|X_{\hat S}$ under $\P$, then for any local maximum $\beta$ of $F(\beta; \P^w)$ 
		over $\R_+^p$, we have 
	\begin{equation*}
		\emptyset \subsetneq \supp(\beta) \subsetneq S \backslash \hat S.
	\end{equation*}	
	\item If $Y|X_S \sim Y|X_{\hat S}$ under $\P$, then $F(\beta; \P^w) = 0$ for all $\beta \in \R_+^p$.
\end{itemize}
\end{proposition}



\begin{proposition}
\label{prop:pop_algo}
Algorithm~\ref{alg:pop} terminates in a set $\hat S$ such that $Y|X \sim Y|X_{\hat S}$ under $\P$ and $\hat S \cap S^c = \emptyset$.
\end{proposition}

The proofs for Propositions \ref{prop:balance}, \ref{prop:reweight_obj} and \ref{prop:pop_algo} are found in 
Appendix \ref{sec:basics-on-algorithm}. Sections~\ref{chap:exclude_noise} and \ref{chap:recovery} consider the statistical properties of the finite sample counterpart of the population algorithm: 
In finite samples, our ability to exclude noise variables is largely preserved, however, the sample complexity needed to recover signal variables will be highly dependent on the type of signal.

\section{The Finite Sample Metric Screening Algorithm}

\label{chap:finite_sample_alg}

In Section \ref{chap:f_restrict}, we studied the population metric screening algorithm and showed that it is variable selection consistent (Proposition \ref{prop:pop_algo}). Next, we study two \emph{finite sample} variants of the metric screening algorithm: Algorithm \ref{alg:metric_screen_low_dim} and Algorithm \ref{alg:metric_screen_high_dim}. The variant algorithms use two distinct mechanisms for controlling false positives, each suited to a different data regime:

\begin{itemize}
  \item In low dimensions ($n \ll p$), Algorithm \ref{alg:metric_screen_low_dim} does not require explicit $\ell_1$ penalization of $\beta$ to exclude noise variables. The metric screening objective is \emph{self-penalizing}: true signals automatically push the coefficients of noise variables to 0.
  \item In high dimensions ($n \gg p$), the self-penalization effect is too weak and Algorithm \ref{alg:metric_screen_high_dim} must employ explicit penalization to induce variable selection. We derive the amount of $\ell_1$ penalization necessary to control false positives.
\end{itemize}

Section \ref{subsec:noise_low_dim} gives the result on false positive control in low dimensions and Section \ref{subsec:noise_result} the result in high dimensions; the complementary discussion of power is deferred to Section \ref{chap:recovery}.

\begin{algorithm}[!tph]
\begin{algorithmic}[1]
\Require{$\gamma >0$, $\alpha > 0$, $\beta^{(0)}$, $\mathbf{X} \in \mathbb{R}^{n\times p}$, $y \in \{0,1\}^n$.}
\Ensure{Initialize $\hat S=\emptyset$, $\Q_n = \sum_{i=1}^n w_i \cdot \delta_{(x_i,y_i)}$ where
	\begin{equation*}
		w_i \propto \begin{cases}
			\#\{y_i=0\}/n~~\text{if}~~y_i = 1 \\
			\#\{y_i=1\}/n~~\text{if}~~y_i = 0
		\end{cases}~~\text{for $i = 1, 2, \ldots, n$}. 
	\end{equation*}
}
\While{$(F(\beta^{(0)}; \Q_n))^2 > \gamma$}
\State{Run projected gradient ascent (with stepsize $\alpha$, initialization $\beta^{(0)}$) to solve 
	\begin{equation*}
		\maximize_{\beta: \beta \in \mathcal{B}} F(\beta; \Q_n).
	\end{equation*}
~~~~Update $\hat S=\hat S \cup \text{supp}(\beta)$ where $\beta$ is any stationary point found by the 
	iterates\footnotemark.}
\State{Estimate $\P(Y|X_{\hat S})$ and update the weighted empirical measure $\Q_n$ by
	\begin{equation*}
		w_i \propto 
			\begin{cases}
				\what{\P}(Y=0|x_{i, \hat S})~~\text{if}~~y_i = 1 \\
				\what{\P}(Y=1|x_{i, \hat S})~~\text{if}~~y_i = 0
			\end{cases}~~\text{for $i = 1, 2, \ldots, n$}.
	\end{equation*}} 
\EndWhile
\end{algorithmic}
\caption{Metric Screening in Low Dimensional Problems}
\label{alg:metric_screen_low_dim}
\end{algorithm}

\footnotetext{\label{note1}Technically, $\beta$ is defined to be any accumulation point of the iterates since 
		there is no prior knowledge that the algorithm will converge to a stationary point. 
		Technically, it is this definition that's used in the proof. }

\begin{algorithm}[!tph]
\begin{algorithmic}[1]
\label{alg:metric_high}
\Require{$\lambda \geq 0$, $\alpha > 0$, $\beta^{(0)}$, $\mathbf{X} \in \mathbb{R}^{n\times p}$, $y \in \{0,1\}^n$.}
\Ensure{Initialize $\hat S=\emptyset$, $\Q_n = \sum_{i=1}^n w_i \cdot \delta_{(x_i,y_i)}$ where
	\begin{equation*}
		w_i \propto \begin{cases}
			\#\{y_i=0\}/n~~\text{if}~~y_i = 1 \\
			\#\{y_i=1\}/n~~\text{if}~~y_i = 0
		\end{cases}~~\text{for $i = 1, 2, \ldots, n$}. 
	\end{equation*}
}
\While{$\hat S$ not converged}
\State{Run projected gradient ascent (with stepsize $\alpha$, initialization $\beta^{(0)}$) to solve 
\begin{equation*}
	\max_{\beta \in \mathcal{B}} F(\beta; \Q_n) - \lambda \cdot |\beta|_1.
\end{equation*}
~~~~~Update $\hat S=\hat S \cup\supp(\beta)$ where $\beta$ is any stationary point found by the iterates\cref{note1}.}
\State{Estimate $\P(Y|X_{\hat S})$ and update the weighted empirical measure $\Q_n$ by
	\begin{equation*}
		w_i \propto \begin{cases}
			\what{\P}(Y=0|x_{i, \hat S})~&\text{if $y_i = 1$} \\
			\what{\P}(Y=1|x_{i, \hat S})~&\text{if $y_i = 0$} 
			\end{cases}~~\text{for $i = 1, 2, \ldots, n$}.
	\end{equation*}}
\EndWhile
\end{algorithmic}
\caption{Metric Screening in High Dimensional Problems}
\label{alg:metric_screen_high_dim}
\end{algorithm}


\subsection{Practical Considerations and Implementation Details}
\label{subsec:implement_details}

For both Algorithm \ref{alg:metric_screen_low_dim} and Algorithm \ref{alg:metric_screen_high_dim}, we maximize a sequence of iteratively reweighted objectives (with optional $\ell_1$ penalty):
\begin{equation}
F(\beta; \Q_n) = \hat{\E}_{n, B-W}[f(\norm{X- X'}_{q, \beta}^q)].
\end{equation}
Projected gradient ascent is used to reach a stationary point of $F(\beta; \Q_n)$ over the constraint set
$\mathcal{B}=\left\{ \beta \in \R_+^p: \norm{\beta}_1 \le  b\right\}$. $\Q_n$ will be a weighted empirical measure with weight $w_i \propto \what{\P}(Y=1-y_i|x_{i, \hat S})$ assigned to datapoint $(x_i, y_i)$. There are two points we wish to emphasize.
\bigskip

\begin{remark}
Our theory (Theorems \ref{prop:false-discovery-low-dim}--\ref{thm:hier_recov}) allows choosing the stepsize to be $\alpha = O(1/p)$ in projected gradient ascent i.e. the stepsize is the same order as the coordinates of the gradient iterates (e.g. $\hat{\beta}^{(0)} = \frac{b}{p} \mathbf{1}$). This 
		implies that with early stopping, gradient ascent can reach an accumulation point with a constant number of iterations\footnote{See the discussion in Section~\ref{sec:remark-on-how-early-stopping-works} and Section~\ref{sec:remark-on-how-early-stopping-works-high} 
in the appendix.}. Here, constant 
		means the number of iterations is independent of $p$. Since each gradient evaluation requires only $O(p)$ flops, a constant number of iterations means that only $O(p)$ total flops are needed to run the entire metric screening algorithm. Thus our claim of an algorithm with linear (in $p$) computation cost is accurate.
\end{remark}

\begin{remark}
To update our weights, we use gradient boosting to estimate $\P(y|x_{\hat S})$ (one is free to choose another nonparametric estimator). We will require
$ \sum_i (y_i - \what{\P}(Y=1|x_{i,\hat S})) = 0$  or equivalently $\Q_n(Y=0) = \Q_n(Y=1)$
which is satisfied by any estimator that includes an unpenalized intercept term. In our proofs, we will ignore estimation error in the weights and simply pretend $\what{\P}(y|x_{\hat S}) = \P(y|x_{\hat S})$. This is not an onerous assumption since $\hat S$ is assumed small i.e. the estimation problem is low dimensional. With some diligent accounting, one can easily adapt our results to allow $\what{\P}(y|x_{\hat S}) \neq \P(y|x_{\hat S})$. Primarily, the thresholds $\gamma$ (Theorem \ref{prop:false-discovery-low-dim}) and $\lambda$ (Theorems \ref{thm:no_false_positive}--\ref{thm:hier_recov}) must be increased to account for additional noise from the estimated weights.
\end{remark}

\section{Controlling False Positives}

\label{chap:exclude_noise}

Our goal in Sections \ref{subsec:noise_low_dim} and \ref{subsec:noise_result} will be to show that the metric screening algorithm successfully excludes noise variables. In both the low and high dimensional regimes, we show that the ability to exclude noise variables applies \emph{simultaneously to all stationary points} of the metric screening objective. It is critical that we are able to provide a guarantee simultaneously for all stationary points since we do not know which stationary point
gradient ascent will bring us to.
Our definition of noise variable will follow Section \ref{subsec:def_noise}.
Let $S$ be the (unique) minimal subset of $\left\{ 1,\ldots,p\right\} $ such that (i) $Y|X\sim Y|X_{S}$
and (ii) $X_{S}\perp X_{S^{c}}$. The set of noise variables is $S^c$.
We make the following assumptions. 
\begin{enumerate}
	\item[(A1)]  $f^\prime$ is strictly completely monotone and $q = 1$ or $q = 2$.
	\item[(A2)]  For some constant $M>0,$ $\norm{X}_\infty \leq M$.
	\item[(A3)]  Imperfect classification: for some constant $\varrho > 0$, 
	\begin{equation*}
		\E\left[\P\left(Y=1|X_{S}\right)|Y=0\right]>\varrho, ~~~
		\E\left[\P\left(Y=0|X_{S}\right)|Y=1\right]>\varrho.
	\end{equation*}
\end{enumerate}
\begin{remark} 
\begin{itemize}
\item (A2) is used primarily for convenience. This assumption can be replaced with
	a subgaussian assumption on the coordinates of $X$. 
\item (A3) is a lower bound on the \emph{effective} sample size. Heuristically, the effective sample size is
\begin{equation*}
\sum_{i: y_{i}=0}w_{i}=\sum_{i: y_{i}=0} \P\big(Y=1|x_{i,\hat{S}}\big)
	\approx n\cdot \E\big[\Var\left(Y|X_{\hat{S}}\big)\right],
\end{equation*}
which is directly proportional to the the amount of \emph{unexplained} variation
in $Y$. The better $X_{\hat{S}}$ is at predicting $Y$ (in our sample),
the harder it becomes to distinguish any remaining signal variables from noise.
This difficulty becomes more pronounced in later rounds of metric screening as $\hat{S}$
becomes bigger or if the class proportions are
heavily unbalanced to begin with.
\end{itemize}
\end{remark}

\subsection{False Positive Control in Low Dimensions}
\label{subsec:noise_low_dim}
In Section \ref{subsec:maineff_ex}, we showed that the population metric screening objective has sparse stationary points that exclude noise variables. Given sufficient sample size, this property is inherited by the empirical metric screening objective! This surprising situation contrasts with, for example, the least squares objective: the minimizer of the population least squares objective excludes noise variables, but the minimizer of the empirical least squares objective has non-zero coefficients for all variables. To induce variable selection while using the least squares objective, one must attach the $\ell_{1}$ or other sparsity inducing penalty. The metric screening objective performs variable selection \emph{without} explicit penalization.

Algorithm \ref{alg:metric_screen_low_dim} controls false positives by checking the termination condition $F^2(\beta^{(0)}; \Q_n) > \gamma$. This condition is a nonparametric test for whether $X \perp Y|X_{\hat S}$. As long as $X$ and $Y$ are \emph{not} conditionally independent, the stationary points of $F_n(\beta; \Q_n)$ will exclude noise variables with high probability (when $X \perp Y$, the population objective is flat). The parallel of Algorithm \ref{alg:metric_screen_low_dim} in the linear model context would be to first conduct an $F$-test for the global null that all coefficients in the linear model are equal to 0 and then minimize the least squares objective if the global null is rejected---in contrast to Algorithm \ref{alg:metric_screen_low_dim}, such a procedure clearly fails to control false positives.

To rigorously state the result, we need two additional assumptions. 
\begin{enumerate}
	\item[(B1)]  Non-degeneracy: for some constant $\zeta > 0$, $\E[|X_j - X_j'|] \ge \zeta$ for $j \not\in S$.
	\item[(B2)] In the case where $q=2$, $f^\prime$ has an analytical 
	extension on the complex plane $\C$ such that $|f^\prime(z)| \le A (1+|z|)^N e^{B |\mathrm{Re}(z)|}$ 
	for some $A, B, N < \infty$.
\end{enumerate}
Both assumptions are mild. (B2) requires 
that when $q=2$, the analytical extension of $f^\prime$ grows at most exponentially on the complex plane 
$\C$\footnote{Many completely monotone function $f^\prime$ satisfies this condition, e.g., $f^\prime(x) = \exp(-x)$.}.

\begin{theorem}
\label{prop:false-discovery-low-dim}
	Assume (A1)-(A3) and (B1)-(B2) hold. Then, there exists a constant $C > 0$ depending only on 
	$f(0), f^\prime(0), f^{\prime\prime}(0), b, M, q, \varrho, \zeta, A,B,N$ such that the following holds:  
	for any $t > 0$, any initialization $\beta^{(0)}$ of the gradient ascent iterates, and any stepsize 
	$\alpha$ and threshold $\gamma$ satisfying 
	\begin{equation*}
	\begin{split}
		\gamma > C \cdot \sqrt{\frac{\log p}{n}} \cdot (1+t) ~~\text{and}~~\alpha \le \frac{1}{C\cdot p},
	\end{split}
	\end{equation*}
with probability at least $1- 2^{s+1}(p^{-t^2} + e^{-n\varrho^2/32})$, the metric learning algorithm 
(Alg.~\ref{alg:metric_screen_low_dim}) outputs $\hat{S} \subseteq S$. 
\end{theorem}

The proof of Theorem~\ref{prop:false-discovery-low-dim} is given in Appendix~\ref{sec:low-dim-recovery}. 

\begin{remark}
	\begin{enumerate}
	\item The fundamental result that lies behind the success of false-discovery control is 
		Proposition~\ref{proposition:null_pop_diff_strengthen} in Appendix~\ref{sec:population-gradient-noise}, 
		which shows that the gradient
		w.r.t any noise variable is upper bounded by the negative of the absolute value of the objective
		(the signal). In other words, the signal automatically penalizes the noise variables.
	\item The check $(F(\beta^{(0)}; \Q_n))^2 > \gamma$ in Algorithm \ref{alg:metric_screen_low_dim} compares the signal against the noise threshold $\gamma$ and ensures that the signal supplies enough implicit regularization to exclude noise variables. This check fails in high dimensions because $F(\beta^{(0)}; \Q_n)$ can vanish exponentially quickly in $p$.
	\end{enumerate}
\end{remark}

\subsection{False Positive Control in High Dimensions}
\label{subsec:noise_result}

In high dimensions, we are still able to guarantee false positive control (simultaneously) for \emph{all}  
stationary points of the metric learning objective. But we must introduce an explicit $\ell_1$ regularization to achieve this.

\begin{theorem}
\label{thm:no_false_positive}
Assume (A1)-(A3) hold. Then, there exists a constant $C > 0$ depending only on 
$f(0), f^\prime(0), f^{\prime\prime}(0), b, M, q, \varrho$ such that the following holds: 
for any $t > 0$, any initialization $\beta^{(0)}$ of the gradient iterates, 
and any stepsize $\alpha$ and penalty $\lambda$ satisfying  
\begin{equation*}
\begin{split}
	\lambda > C \cdot \sqrt{\frac{\log p}{n}} \cdot (1+t)~~\text{and}~~\alpha \le \frac{1}{C \cdot p},
\end{split}
\end{equation*}
the metric learning algorithm (Alg.~\ref{alg:metric_screen_high_dim}), with probability at least 
$1- 2^{s+1}(p^{-t^2} + e^{-n\varrho^2/32})$, excludes all noise variables---its output $\hat{S}$ satisfies 
$\hat{S} \subseteq S$. 
\end{theorem}
The proof of Theorem~\ref{thm:no_false_positive} can be found in Appendix~\ref{sec:high-dim-false-discovery}. Theorem \ref{thm:no_false_positive} follows from our characterization of the landscape of the 
(non-convex) metric learning objective---Proposition~\ref{prop:no-stationary-point-A} in Appendix~\ref{sec:high-dim-false-discovery} 
shows 
that the coefficient of $\beta_{j}$ for any noise variable will be $0$ (w.h.p.) for \emph{any stationary point} of the 
penalized objective $F(\beta; \Q_n) - \lambda \cdot \norm{\beta}_1$.



\section{Recovering Signals}

\label{chap:recovery}

We analyze the power of the metric screening algorithm. The proofs of our results all follow a common logic: show that gradient ascent carries us to a non-zero stationary point of the metric learning objective. Theorems \ref{prop:false-discovery-low-dim} and \ref{thm:no_false_positive} tell us that whenever we reach a non-zero stationary point, we will have recovered true (and only true) variables. The argument for \emph{why} we are able to reach a non-zero stationary point is different in low and high dimensions.

In low dimensions, almost any initialization, $\beta^{(0)}$, is a \emph{good} initialization. By good, we mean that the objective value at $\beta^{(0)}$ exceeds the noise threshold, $\gamma$, and also exceeds 0. Since gradient ascent monotonically increases the objective value, a good initialization implies that we will reach a non-zero stationary point. Formalizing this argument, we provide a quite general result (Theorem \ref{thm:recov-low-dim}), guaranteeing recovery of all non-redundant variables in low dimensions. This result holds for both $\ell_1$ ($q=1$) and $\ell_2$ ($q=2$) based metric learning objectives.

Unfortunately, the probability of stumbling on a good initialization without prior knowledge vanishes exponentially quickly in $p$ (Section \ref{subsec:gen_pure_eff}). In high dimensions, the argument for why we reach a non-zero stationary point relies on characterizations of gradient dynamics which we are able to do only for specific classes of signals. We show that:
\begin{itemize}
		\item We can recover main effects with $n\sim \log p$ samples.
		\item We can recover pure order-$s$ interactions with $n \sim p^{2(s-1)}\log p$ samples (when constrained by linear computation budget).
		\item We can recover hierarchical interactions with $n \sim \log p$ samples.
\end{itemize}
Surprisingly, our high dimensional recovery results can be attained if we use $\ell_1$ distance ($q=1$) in the metric learning objective but \emph{not} if we use $\ell_2$ distance ($q=2$). A seemingly accessory aspect of the metric learning objective turns out to have an outsize impact on performance! We exhibit signals where the metric learning objective has arbitrarily poor detection power if $q=2$.  In contrast, Section \ref{subsec:gen_pure_eff} shows that when $q=1$, one can make certain guarantees on the gradients of the metric learning objective which hold for \emph{all} signal types.

\subsection{Recovery in Low Dimensions}

Given sufficient signal strength, metric screening will return a set $\hat S$ such that $Y|X_{\hat S} \sim Y|X_S$. To prove this, we show that Algorithm \ref{alg:metric_screen_low_dim} cannot terminate while $Y|X_{\hat S} \neq Y|X_S$, or equivalently, that the check $(F(\beta^{(0)}; \Q_n))^2 \ge \gamma$ keeps passing while there is signal yet to be discovered. Note that $F(\beta^{(0)}; \Q_n)$ vanishes exponentially quickly in $p$; but in low dimensions, we can treat it as $O(1)$.

\label{subsec:recov-low-dim}
\begin{theorem}
\label{thm:recov-low-dim}
	Make the same assumptions and use the same threshold $\gamma$ and stepsize $\alpha$ as in 
	Theorem \ref{prop:false-discovery-low-dim}. Then with probability at least $1- 2^{s+1}(p^{-t^2} + e^{-n\varrho^2/32})$, 	
	the metric learning algorithm (Alg.~\ref{alg:metric_screen_low_dim}) outputs $\hat{S}$ satisfying 
	$Y \mid X_{\hat{S}} = Y\mid X_S$ if the following condition holds: 
	\begin{equation}
	\label{eqn:condition-in-low-dimension}
		\inf_{A: A \subseteq S, Y\mid X_A \neq Y\mid X_S}
			\left(F(\beta^{(0)}; \Q^A)\right)^2 \ge 2\gamma.
	\end{equation}
	In above, $\Q_A$ is the probability distribution satisfying $\frac{d\Q^A}{d\P} (x, y) \propto w_A(x, y)$
	where $w_{A}(x,y)=1-\P(Y=y|X_{A}=x_{A})$ and $\P$ is the true distribution of the raw data $(X,Y)$.

	
\end{theorem}

\begin{remark}
$F(\beta^{(0)}; \Q^A)$ quantifies the remaining signal after removing the effect of $X_A$. Condition \ref{eqn:condition-in-low-dimension} says that for all subsets $A\subsetneq S$ such that $Y|X_A \neq Y|X_S$, the gap in predictive power between $X_S$ and $X_A$ is large relative to the noise threshold $\gamma = \Omega(\sqrt{\log p/n})$.
\end{remark}

\subsection{Recovery of Main Effects}

\label{subsec:maineff_recov}
Let $S= [s]$ denote the set of signal variables and assume that the signal variables are conditionally independent: $X_{1}\perp\ldots\perp X_{s}|Y$. The discriminative model $Y|X$ becomes an additive logistic model and the conditional independence assumption means that the signal variables have \emph{orthogonal} main effects, $f_j(X_j)$, in the additive model:
\begin{equation*}
\text{logit}~\P\left(Y=1|X_A\right)=\alpha+\sum_{j \in [s] \cap A} f_{j}\left(X_j\right), \qquad \forall A\subseteq [p].
\end{equation*}
Given orthogonal main effects, we show that $\hat S = S$ with $n\sim \log p$ samples. The key property of the main effects model which allows for this recovery result is that w.h.p. $\partialbeta F_n(0) > c > 0$ for all variables 
$j \in S$. Hence, $\beta=0$ is never a stationary point.

\begin{theorem}
\label{thm:main_eff_recov}
Assume (A1)-(A3) hold. Let $q = 1$. 
There exists some constant $C > 0$ depending only on $f(0), f^\prime(0), f^{\prime\prime}(0), M, b, \varrho$ 
such that the following holds: for any $t > 0$, and any initialization $\beta^{(0)}$ of the gradient iterates, 
any stepsize $\alpha$ and any threshold $\lambda> 0$ satisfying 
\begin{equation*}
\lambda > C \cdot \sqrt{\frac{\log p}{n}} \cdot (1+t)~~\text{and}~~\alpha \le \frac{1}{C \cdot p},
\end{equation*}
with probability at least $1- 2^{s+2} (p^{-t^2} + e^{-n\varrho^2/32})$, the metric learning 
(Alg.~\ref{alg:metric_screen_high_dim}) returns 
$\hat{S}$ satisfying $S \supseteq \hat{S} \supseteq S(\lambda)$, where 
$S(\lambda) \subseteq S$ is defined by
\begin{equation*}
\begin{split}
S(\lambda) \defeq \Big\{j \in S: \Signal(\{j\}) \ge 2\lambda \Big\},~~~~~~~~~~~~ \\
~~\text{where}~~\Signal(\{j\}) = f^\prime(0) \cdot \popexpdiff\left[\left|X_{j}-X_{j}'\right|\right]~\text{for $j \in S$}. 
\end{split}
\end{equation*}
\end{theorem}

The proof of Theorem~\ref{thm:main_eff_recov} can be found in Appendix~\ref{sec:proofs_recovery}.

\begin{remark}
We use the conditional independence between signal variables to rule out redundant variables
i.e. variable $j$ contributes little additional explanatory power
after accounting for the other variables. The conditional independence assumption is overly strong for this purpose. Proposition \ref{cor:dependent_main} in Appendix~\ref{sec:proofs_all_recovery} proves recovery of main effects 
by directly assuming that $X_{j}$ has an effect on $Y$ conditional
on $X_{A}$ for any subset $A\subseteq S \backslash \left\{ j\right\} $, thus bypassing the need for conditional independence.
\end{remark}

\begin{remark}
To quantify the effect size of a main effect signal $j$, Theorem \ref{thm:main_eff_recov} uses the quantity 
$\Signal(\{j\}) = f^\prime(0) \cdot \popexpdiff\big[|X_{j}-X_{j}'|\big]$. 
This is justified by the fact that $X_{j}\perp Y$ if and only if $\Signal(\{j\})=0$.
\end{remark}

Note that $n\sim\log p$ is also the sample size needed to recover variables in a linear logistic
model using the lasso. The advantage of metric screening is that we can detect nonlinearities
(e.g. unequal variances across class which manifests as a quadratic
term in a logistic model) without having to prespecify the form of the nonlinearities.


\subsection{Recovery of Pure Interactions}

\label{subsec:pure_recov}

We show that given a computational budget of $p^{s_0}$, we can recover a \emph{pure} interaction with $n\sim p^{2\left(s-s_0\right)_+}\log p$ samples.  Under a pure interaction signal, the
landscape of the population metric loss objective has a couple of distinctive features:
\begin{itemize}
\item All non-zero stationary points $\beta$ are supported on $S$.
\item $\beta = 0$ is always a stationary point (not the case for main effects!).
\end{itemize}
The second point makes recovery of pure interactions particularly difficult in high dimensions. In the absence of prior information, 
most reasonable initializations for $\beta$ will have small coefficients for all variables and will tend to fall into the basin of attraction around 
$\beta = 0$. To recover a pure interaction, the path taken by gradient ascent must roughly have the following property: in any iteration,
\emph{no more than one} of the $\beta_{j}$ for $j\in S$ can be ``small''. In Theorem \ref{thm:pure_recovery}, we show that when 
initializing at $\beta^{(0)} = \frac{b}{p} \mathbf{1_p}$, every gradient ascent step $k$ maintains $\beta_{j}^{(k)}\gtrsim \frac{1}{p}$
for all $j\in S$.

\begin{theorem}
\label{thm:pure_recovery}
Assume (A1)-(A3) hold. Let $q = 1$. Assume $X_S$ is a pure interaction, i.e., $X_A \perp Y$ for all 
strict subset $A \subsetneq S$. 
Then there exists some constant $C > 0$ depending only on $f(0), f^\prime(0), f^{\prime\prime}(0), M, b, \varrho$
such that the following holds: for any $t > 0$, any stepsize $\alpha$, 
and threshold $\lambda$ satisfying
\begin{equation*}
\lambda > C \cdot \sqrt{\frac{\log p}{n}} \cdot (1+t)~~\text{and}~~
\alpha \le \frac{1}{C (p\cdot s)}, 
\end{equation*}
with probability at least $1-(p^{-t^2} + e^{-n\varrho^2/32})$, the metric learning (Alg.~\ref{alg:metric_screen_high_dim}) with 
initialization $\beta^{(0)} = \frac{b}{p} \mathbf{1_p}$ returns $\hat{S} = S$ if $\Signal(S) > 2\lambda + \frac{C}{p^s}$ where
\begin{equation*}
\begin{split}
\Signal(S) = 
\frac{1}{8^s b} \cdot \left|\frac{f^{(s)}(Mb)}{f^{(s)}(0)}\right| \cdot \left(\frac{1}{p}\right)^{s-1} \cdot 
F(b\mathbf{1}_S; \P). 
\end{split}
\end{equation*}
\end{theorem}

The proof of Theorem~\ref{thm:pure_recovery} can be found in Appendix~\ref{sec:proof_pure}. The sample size requirement, $n\sim p^{2\left(s-1\right)}\log p$,
for the recovery of a pure order $s$ interaction is quite high. We
emphasize though that this is the sample size requirement when we
restrict ourselves to a computation budget that is \emph{linear} in the number of variables.
With a larger computational budget, we can increase our chances of reaching a non-zero stationary point
by trying multiple initializations of $\beta$. Assuming a computation budget of 
$p^{s_{0}}$ (i.e. try $O(p^{s_0 - 1})$ initializations), we can detect an order $s$ pure interaction w.h.p.  
with $n\sim p^{2\left(s-s_{0}\right)}\log p$ samples. For details, see 
Section~\ref{sec:discussion-on-reducing-computational-complexity} in 
Appendix~\ref{sec:proofs_all_recovery}.

\subsection{Recovery of Hierarchical Interactions}

\label{subsec:hierint_example}

Many popular methods focus exclusively on finding hierarchical interactions. A potential concern about metric screening is that it might pay for its ability to detect pure interactions by performing much worse with respect to hierarchical interactions. We ally this fear by showing that the metric screening algorithm automatically adapts when the signal is hierarchical, finding such interactions with only $n\sim\log p$ samples in contrast to the $n \sim p^{2(s-1)} \log p$ samples needed for pure interactions.

The notion of a hierarchical interaction
has traditionally been defined with respect to specific parametric
models and is difficult to extend to the nonparametric
setting. We will adopt a very
stylized definition of hierarchical interaction. Our goal here is merely to provide a fable illustrating \emph{how} metric screening exploits the existence of main
effects to find hierarchical interactions more efficiently than pure interactions.


\begin{definition}[Hierarchical Interaction]
The variables in $S$ interact hierarchically
if there exists a nested sequence of variables $\emptyset=S_{0}\subsetneq S_{1}\subsetneq S_{2}\subsetneq \ldots\subsetneq S_{s} = S$
with $\left|S_{k}\right|=k$ such that for all $1\leq k\leq s$,
\begin{itemize}
\item $X_{S_{k} \backslash S_{k-1}}$ is not independent of $Y$ given $X_{S_{k-1}}$.
\item $X_{S \backslash S_{k}}\perp Y|X_{A}$ for any strict subset $A\subsetneq S_{k}$.
\end{itemize}
\end{definition}

Given the existence of main effects in hierarchical interactions, one may wonder why Theorem
\ref{thm:main_eff_recov} or Proposition \ref{cor:dependent_main} (Appendix \ref{sec:proofs_all_recovery}) do not automatically imply their recovery.
Suppose $X_1$ and $X_2$ are involved in a hierarchical interaction and $X_1$ has a main effect on $Y$. 
The situation we must account for is when variable 1 (but not variable 2) is selected in the first round of 
metric screening and rebalancing with respect to variable 1 leads to a reweighted distribution where 
$X_1$ and $X_2$ are now involved in a pure interaction. Example \ref{ex:hier_to_pure} in Appendix \ref{subsec:hier_example} shows 
that the conversion of a hierarchical interaction into a pure interaction (after rebalancing) is not a farfetched 
situation. To deal with Example \ref{ex:hier_to_pure} and related cases, we'll need to modify the metric 
screening algorithm to take advantage of the nested structure of a hierarchical interaction. In particular, 
after a variable has been selected, we will initialize it at $\beta_{k}= \tau>0$ in subsequent rounds of metric screening 
and minimizing the metric screening objective subject to $\beta_{k}= \tau$. This modified metric screening algorithm recovers hierarchical interactions with $n \sim \log p$ samples (see Algorithm \ref{alg:metric_screen_hier} in Section~\ref{sec:proof_hier} in Appendix~\ref{sec:proofs_all_recovery}).


\begin{theorem}
\label{thm:hier_recov}
Assume (A1)-(A3). Let $q = 1$. 
Assume $X_S$ is a hierarchical interaction with the nested sequence 
$\emptyset \subsetneq S_1 \subsetneq S_2 \cdots \subsetneq  S_{s}= S$
with $|S_k| = k$. There exists a constant $C > 0$ depending 
only on $f(0), f^\prime(0), f^{\prime\prime}(0), M, b, \varrho$ such that the following holds: 
for any $t > 0$, any initialization $\beta^{(0)}$ of the projected gradient iterates, 
any stepsize $\alpha$ and any threshold $\lambda> 0$ satisfying 
\begin{equation*}
\lambda > C \cdot \sqrt{\frac{\log p}{n}} \cdot (1+t)~~\text{and}~~\alpha \le \frac{1}{C \cdot p},
\end{equation*}
with probability at least $1-2^{s+2}(p^{-t^2} + e^{-n\varrho^2/32})$, the metric learning algorithm
(Alg.~\ref{alg:metric_screen_hier}) returns a set $\hat{S}$ that satisfies 
$S \supseteq \hat{S} \supseteq S_{l(\lambda, \tau)}$ where $l(\lambda, \tau)$ is the largest 
integer $l \in [s]$ such that for all $1\le k \le l$, 
\begin{equation*}
\begin{split}
\min_{A: S_{k}\subseteq A}\Signal(A \mid S_{k-1}; \tau) \ge 2\lambda,~~~~~~~~~~~~~~~~~~~~~~~	 \\
~~\text{where}~~\Signal(A \mid S_{k-1}; \tau) 
	=  \frac{1}{\tau}\cdot \popexpdiff^{({S_{k-1}})}
		\left[f(\tau \cdot \norm{X_{A}-X_{A}'}_1)\right]. 
\end{split}
\end{equation*}
\end{theorem}

The proof of Theorem~\ref{thm:hier_recov} can be found in Appendix~\ref{sec:proof_hier}. 

\subsection{The Importance of Choosing $q=1$}

\label{subsec:gen_pure_eff}


In this subsection, we give an overview of why $\ell_1$ kernels have better variable selection properties. 
Rigorous statements of the special properties of the $\ell_1$ kernel (and their proofs) are deferred to 
Appendix~\ref{sec:proofs_laplace}; these properties form the foundation for our recovery proofs. 
To appreciate the power of $\ell_1$ kernels,  we first consider a very simple example. 

\bigskip

\newcommand{\gauss}{{\rm G}}
\newcommand{\Lap}{{\rm L}}
\begin{example*}
\label{ex:unequal_variance}
Consider the following model where we have only one signal variable $X_1$ (and $X_2, \ldots, X_p$ are pure noise): 
\begin{equation*}
\begin{split}
	X_{1}|Y=0\sim \normal\left(0,\sigma_{0}^{2}\right),~~X_{1}|Y=1\sim \normal\left(0,\sigma_{1}^{2}\right), \\
	X_{j}|Y= 0 \sim X_j \mid Y = 1 \sim \normal\left(0, \sigma^2 \right)~\text{for $j \neq 1$},~~ \\
	X_1 \perp X_2 \perp \ldots \perp X_p \mid Y,~~~~~~~~~~~~~~~~~
\end{split}
\end{equation*}
where $\sigma_{0}^{2}= \sigma^2(1 - \delta)$ and $\sigma_{1}^{2}= \sigma^2 (1 + \delta)$
for some $\delta \in (0, 1)$. This model assumes that $X_{1}$ has a main effect on $Y$ in the 
form of unequal variances across class. We consider two variants of the metric learning objective: 
the first, denoted $F^{\gauss}$, corresponds to the metric learning objective using the Gaussian kernel 
($f(x)=-e^{-x}$ and $q=2$) and the second, $F^{\Lap}$, corresponds to the 
Laplace kernel ($f(x)=-e^{-x}$ and $q=1$). 

Suppose we initialize gradient ascent at $\beta^{(0)} = \frac{1}{p} \mathbf{1}_p$. A simple calculation shows that 
\begin{equation}
\begin{split}
	F^{\gauss}(\beta^{(0)}) \sim p^{-2}~~&\text{and}~~ F^{\Lap}(\beta^{(0)}) \sim p^{-1}, \\
	\frac{\partial}{\partial \beta_1}F^{\gauss}(\beta^{(0)}) \sim p^{-1}~~&\text{and}~~
	\frac{\partial}{\partial \beta_1}F^{\Lap}(\beta^{(0)}) \sim 1. 
\end{split}
\end{equation}
The choice of $q$ impacts how quickly the signal vanishes as dimensionality increases. The gradient for the Laplace kernel ($q=1$) does not diminish with increasing 
dimensionality---even in high dimensions, gradient ascent will tend to increase the value of the coefficient for $X_1$, helping bring us to a non-zero stationary point. On the other hand, finite sample noise (and the $\ell_1$ penalty we introduce to control the noise) will easily trump the gradient for the Gaussian kernel ($q=2$) when $p$ is large.
\end{example*}

The superiority of $q=1$ in the above example is 
not coincidental and holds for all reasonable choices of $f$. Proposition~\ref{proposition:l_1-separates-from-l_2} exhibits the general phenomenon: we construct examples where the gradient of the metric learning objective (w.r.t. signal variables) is guaranteed to be $\Omega(p^{-|S|+1})$ when $q=1$ but is only $O(p^{-l})$ for arbitrarily large $l \in \N$ when $q=2$.

\begin{proposition}
\label{proposition:l_1-separates-from-l_2}
Let $f \in \mathcal{C}^\infty(\R_+)$ be such that $f^\prime$ is strictly completely monotone. Fix $0 < t \le b$ and 
$\bar{\beta} = \frac{t}{p} \mathbf{1}$. Fix the distribution $(X_S, Y) \sim \P_0$ where $\P_0$ is a pure-interaction. 
Consider the set of distributions on $\R^{p+1}$: 
\begin{equation*}
	\mathcal{Q}(p, M, S, \P_0) = \left\{\Q: 
		\norm{X}_\infty \le M~\text{and}~ (X_S, Y) \sim \P_0~\text{for $(X, Y) \sim \Q$}\right\}.
\end{equation*}
That is $\mathcal{Q}(p, M, S, \P_0)$ is the set of distribution on $(X, Y)$ who has compact support 
and whose marginal distribution satisfies $(X_S, Y) \sim \P_0$.
\begin{itemize}
\item Assume $q = 1$. Fix the distribution $\P_0$. Then for any sequence of distributions  
	$\{\Q_p\}_{p \in \N}$ where $\Q_p \in \mathcal{Q}(p, M, S, \P_0)$, we have the lower bound 
	\begin{equation}
	\label{eqn:l_1_signals-prop}
	\begin{split}
		F \left(\bar{\beta}; \Q_p, f, 1\right) &= \Omega (p^{-|S|}), \\
		\frac{\partial}{\partial \beta_j}F \left(\bar{\beta}; \Q_p, f, 1\right) &= \Omega(p^{-(|S|-1)})~\text{for any $j \in S$}.
	\end{split}
	\end{equation}
\item Assume $q = 2$. Fix any $l \in \N$. There exists a pure-interaction $\P_0$ such that any sequence 
	$\{\Q_p\}_{p \in \N}$ where $\Q_p \in \mathcal{P}(p, M, S, \P_0)$ must satisfy
	\begin{equation}
	\label{eqn:l_2_no-signals-prop}
	\begin{split}
		F \left(\bar{\beta}; \Q_p, f, 2\right) &= O (p^{-l}), \\
		\frac{\partial}{\partial \beta_j}F \left(\bar{\beta}; \Q_p, f, 2\right) &= O(p^{-l})~\text{for any $j \in S$}.
	\end{split}
	\end{equation}
\end{itemize}
The way to appreciate equations~\eqref{eqn:l_1_signals-prop}-\eqref{eqn:l_2_no-signals-prop} is 
to view both sides of equations as a function of $p$, with the rest of the parameters $f, M, b, S, t, \P_0$ fixed.
\end{proposition}

The proof of Proposition~\ref{proposition:l_1-separates-from-l_2} can be found in Appendix~\ref{sec:proof-proposition:l_1-separate-from-l_2}. The proof relies on two key properties which hold for all $\ell_1$ type kernels (see Appendix~\ref{sec:proofs_laplace} 
for formal statement and proof): 
\begin{itemize}
\item Lemma \ref{lem:deriv_bound} relates $\partialbeta F\left(\beta\right)$
to $F\left(\beta\right)$: roughly, 
\begin{equation*}
	\partialbeta F\left(\beta\right) \approx \beta_j^{-1} F(\beta)~~\text{for any $j \in S$}.
\end{equation*} 
The gradient vanishes slower than the objective value as $\beta \rightarrow 0$.
\item Lemma \ref{lem:self_bounding} bounds how quickly $F\left(\beta\right)$ can vanish as $\beta_j$ becomes small: for any signal, $F\left(\beta\right)$ decreases at worst linearly in $\beta_{j}$ for $j \in S$.
\end{itemize}
These properties of $\ell_1$ kernels derive from the fact that their Fourier transform is heavy tailed (see 
Appendix \ref{sec:proofs_laplace}). In contrast to the $\ell_1$ kernel, the Fourier transform of the $\ell_2$ 
kernel is light tailed with an exponential decay at $\infty$. This is the underlying reason why 
$\ell_2$ kernels do not enjoy the nice properties of the $\ell_1$ kernel.

\section{Experimental Results}

\label{chap:simulatons}

We empirically compare the metric screening algorithm against a trio of competitors: (i)
logistic regression with a lasso penalty, (ii) marginal screening via the distance correlation statistic and (iii) random forest variable importance. Logistic regression assumes a linear signal. Distance correlation allows us to capture nonlinear signals but used in conjunction with marginal screening means that we may miss interactions. Random forest allows us to capture nonlinear signals and hierarchical interactions but may miss pure interactions. Finally, metric screening has detection power against all signals but pays an efficiency cost of course for its generality. We will test the methods on four simulated examples: one with main effects only signal, two with hierarchical interactions, and one with pure interactions.

In the simulations, we use each method to choose $|S|$ variables and evaluate the fraction of true variables that were recovered. In practice, one would not have access to the number of true variables $|S|$. This element of our experiment design was necessary to create a fair comparison: tree based variable importance and marginal
screening provide only a ranking of the variables and do not specify
the number of variables the user should keep, so we fix in advance the number of variables chosen at $|S|$. For
the lasso, we choose the $\ell_{1}$ regularization
so that exactly $|S|$ variables are chosen; for metric screening, we terminate Algorithm \ref{alg:metric_screen_low_dim} once $|S|$ variables have been chosen.\footnote{Ties are broken using the size of the
corresponding coefficients.}

\subsection{Main Effects Only}

\begin{figure}[!tph]
\begin{centering}
\includegraphics[width=.5\linewidth]{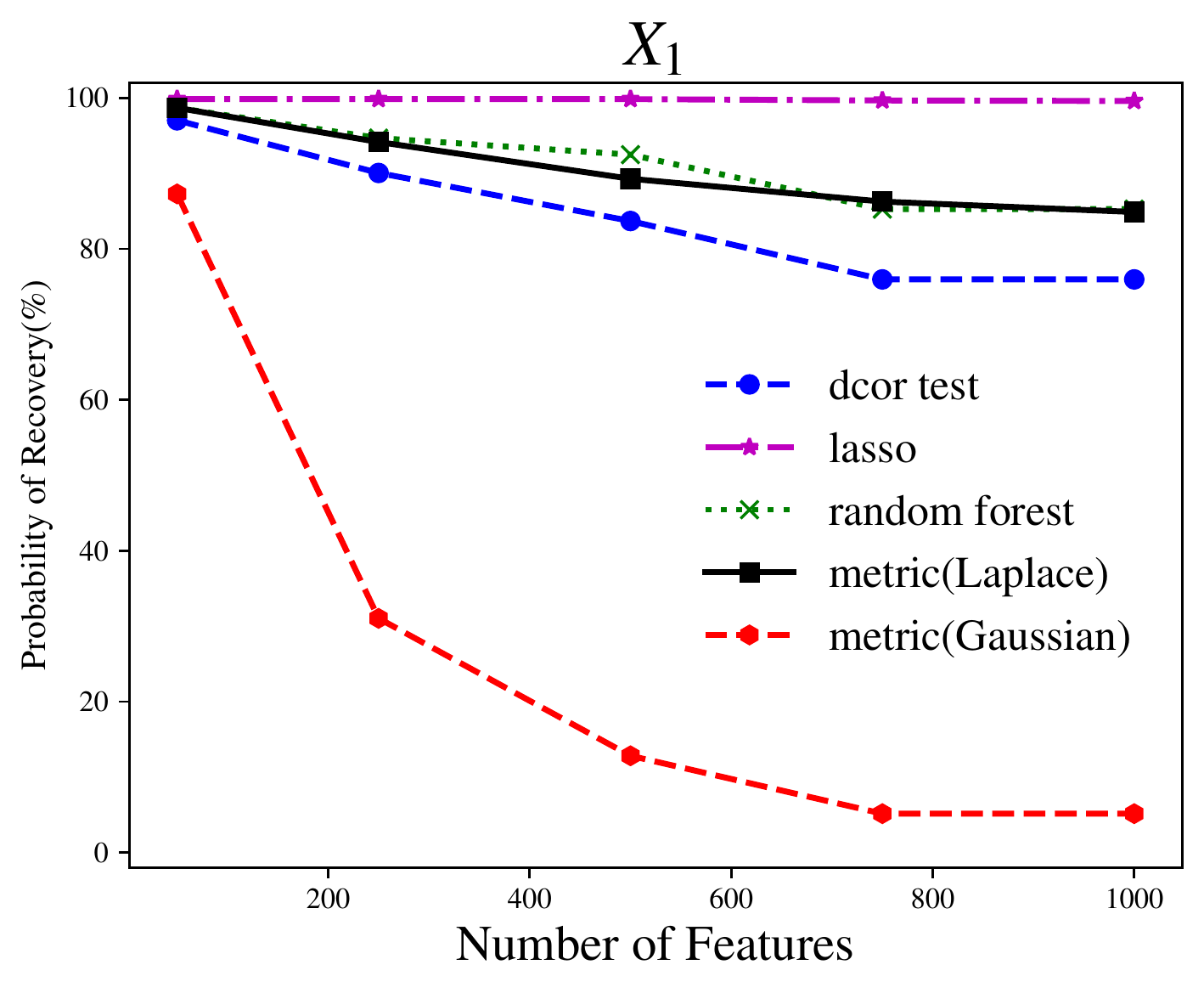}\includegraphics[width=.5\linewidth]{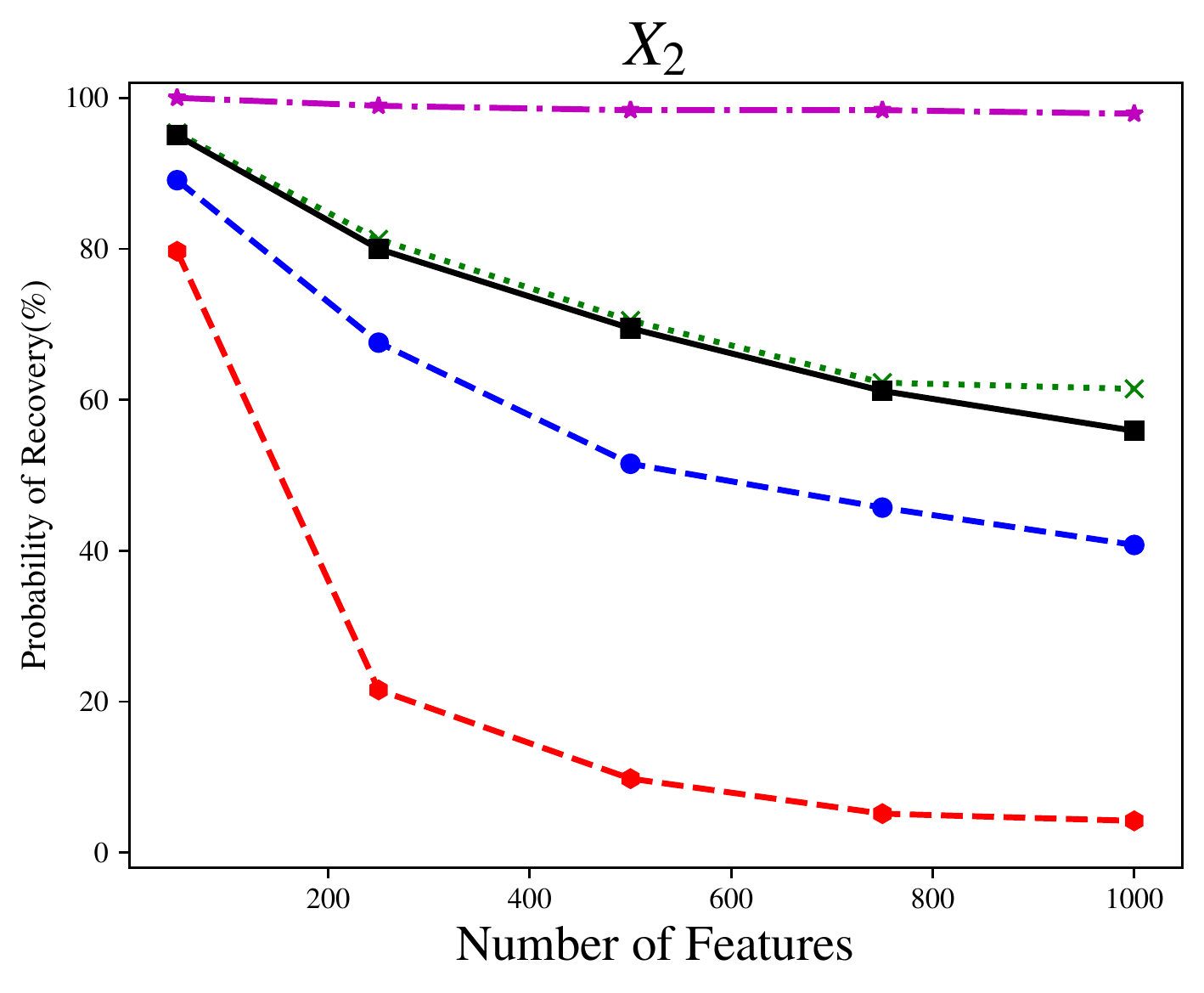}
\par\end{centering}
\begin{centering}
\includegraphics[width=.5\linewidth]{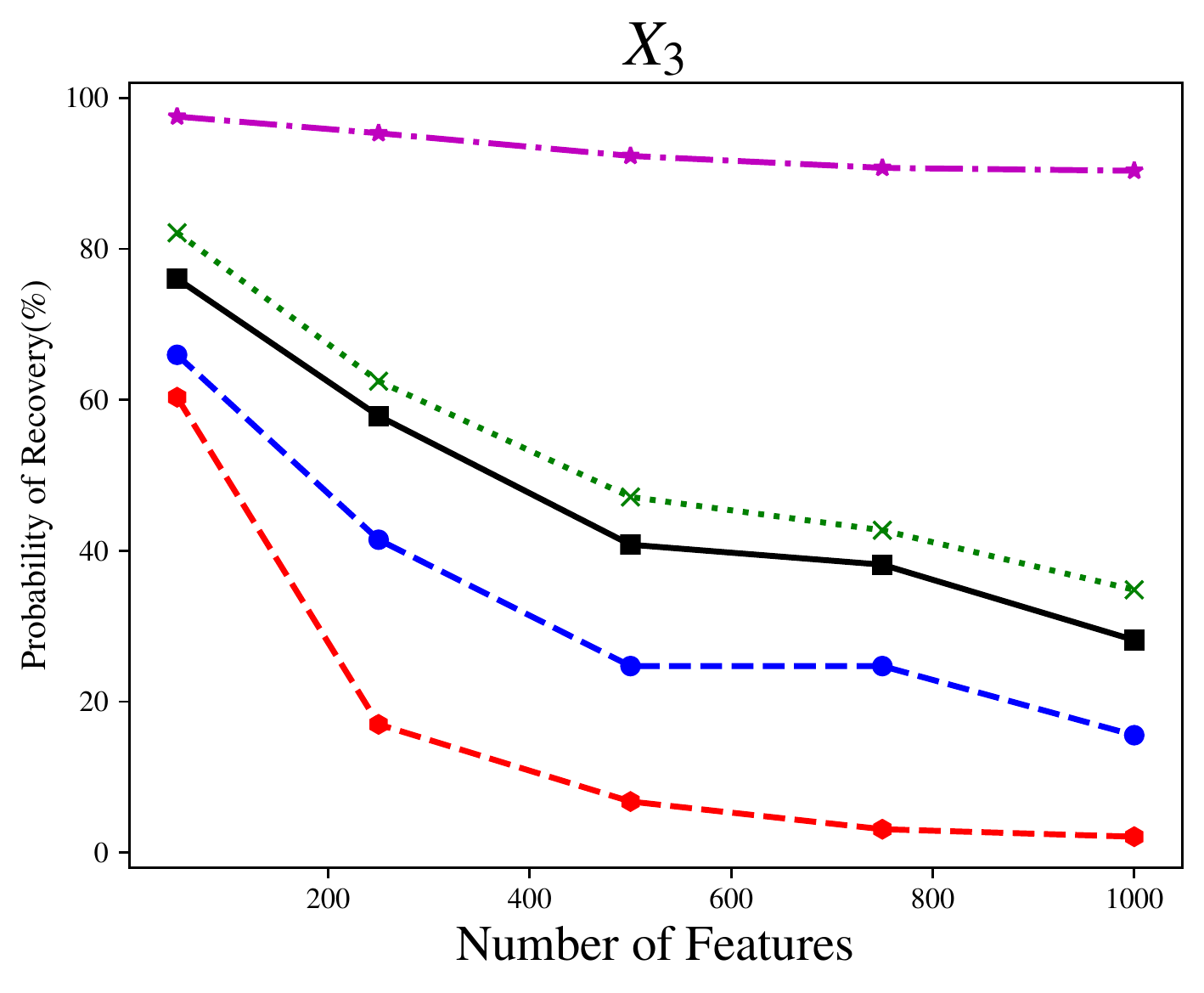}\includegraphics[width=.5\linewidth]{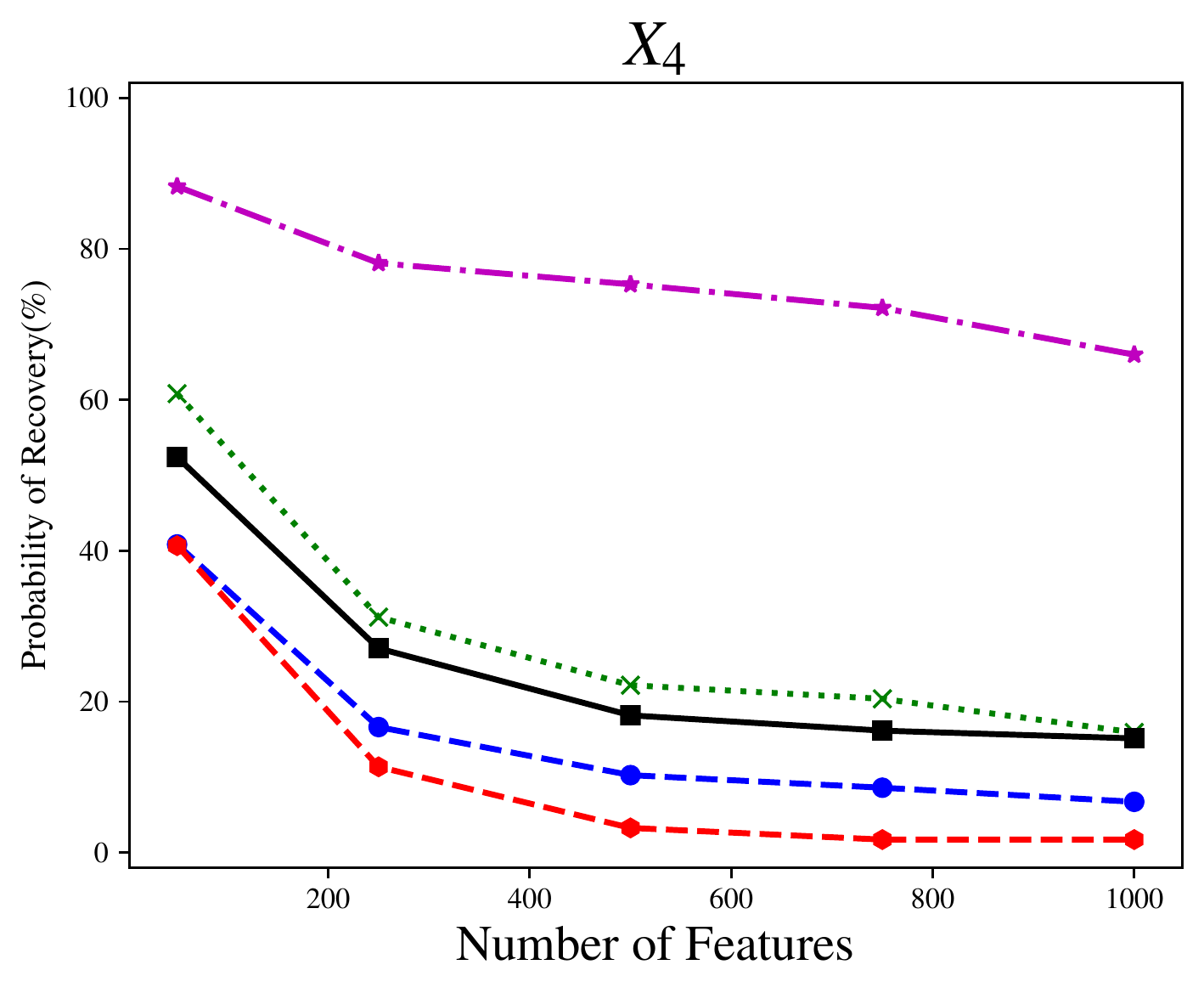}
\par\end{centering}
\caption{\small Probability of recovering true variables in an unequal variance
model: $X_{j}|Y=1\sim \normal\left(0, \sigma^2 (1+\delta_j)\right)$
and $X_{j}|Y=0\sim \normal\left(0, \sigma^2 (1-\delta_j)\right)$
for $j=1,\ldots,4$ (true variables) and $X_{j}|Y\sim \normal\left(0,\sigma^{2}\right)$
for $j>4$ (null variables). Here $\sigma^{2}=1$ and $\left(\delta_{1},\delta_{2},\delta_{3},\delta_{4}\right)=\left(0.4,0.35,0.3,0.25\right)$.
The sample size $n=500$ is fixed while the number of null variables
ranges over $\left\{ 50,250,500,750,1000\right\} $.}

\label{fig:uneq_var}
\end{figure}

We use the model of unequal class variances described in Example \ref{ex:unequal_variance}:
\begin{equation*}
\begin{split}
&\begin{split}
X_j \mid Y = 0 \sim \normal \left(0, \sigma^2 (1+\delta_j)\right) \\
X_j \mid Y = 1 \sim \normal \left(0, \sigma^2 (1-\delta_j)\right)
\end{split}~~\text{for $j = 1, \ldots, 4$}, \\
&X_j \mid Y = 0 \sim X_j \mid Y = 1 \sim \normal(0, \sigma^2)~~\text{for $j > 4$}.
\end{split}
\end{equation*}
There are 4 true variables and the remaining
variables $X_{j}$ for $j>4$ are noise. We set $\sigma^{2}=1$ and
$\left(\delta_{1},\delta_{2},\delta_{3},\delta_{4}\right)=\left(0.4,0.35,0.3,0.25\right)$
so that both weak and strong predictors are present. The discriminative
model of $Y|X$ is a logistic model linear in the quadratic terms
$X_{1}^{2},$ $X_{2}^{2}$, $X_{3}^{2}$, and $X_{4}^{2}$. When running
the lasso, we fed the algorithm the transformed predictors, $X_{j}^{2}$,
rather than the raw predictors, $X_{j}$; hence the lasso results
for this experiment can be interpreted as an oracle benchmark.
The simulation results are plotted in Figure \ref{fig:uneq_var}.

We compared two variants of metric screening, one using the Gaussian
kernel ($q=2$) and one using the Laplace kernel ($q=1$). The Laplace
kernel significantly outperforms the Gaussian kernel in recovering
the true variables. This validates the theoretical analysis presented in Example
\ref{ex:unequal_variance}. Of the three nonparametric methods, random
forest variable importance is the winner, slightly outperforming metric screening (Laplace). Surprisingly, despite the absence of interaction signals, distance correlation based marginal screening performed the worst.

\subsection{Hierarchical Interactions: QDA Model}

We consider a QDA model with four true variables
\begin{equation}
\label{eq:qda_model}
\begin{split}
(X_1, \ldots, X_4) \mid Y = 1 \sim \normal(+\mu, \Sigma^+) \\
(X_1, \ldots, X_4) \mid Y = 0 \sim \normal(-\mu, \Sigma^-)
\end{split},
\end{equation}
where
\begin{equation*}
\mu = (\delta_1, \xi, \delta_2, \xi),~~
\Sigma^+ = 
\begin{pmatrix}
1 & +\rho \\
+\rho & 1 
\end{pmatrix},~~
\Sigma^- = 
\begin{pmatrix}
1 & -\rho \\
-\rho & 1 
\end{pmatrix}.
\end{equation*}
In above, $\delta_{1}$ and $\delta_{2}$ govern the main effects of $X_{1}$
and $X_{3}$ respectively; $\xi$ governs the main effects of $X_{2}$
and $X_{4}$; $\rho$ governs how strongly $X_{1}$ interacts with
$X_{2}$ and $X_{3}$ with $X_{4}$. We set $\delta_{1}=0.25$, $\delta_{2}=0.2$, $\xi=0.1$, and $\rho=0.5$. Since $\xi$ is small, $X_2$ and $X_4$ will be hard to detect if one solely relies on their marginal relationship with $Y$. There are four true variables. Additional noise variables are generated
as independent standard normals: $X_{j}|Y\sim \normal\left(0,1\right)$
for $j>4$. The simulation results are plotted in Figure \ref{fig:qda}. Neither the lasso nor distance correlation based marginal screening can
detect interactions. These two methods have no problem selecting $X_{1}$
and $X_{3}$ but fare poorly in
detecting $X_{2}$ and $X_{4}$. In low dimensions, both random forest and metric screening do a good job finding $X_{2}$ and $X_{4}$. In higher dimensions, metric screening fares better.

\subsection{Hierarchical Interactions: Ratio Based Signal}
\begin{figure}[!tph]
\begin{centering}
\includegraphics[width=.5\linewidth]{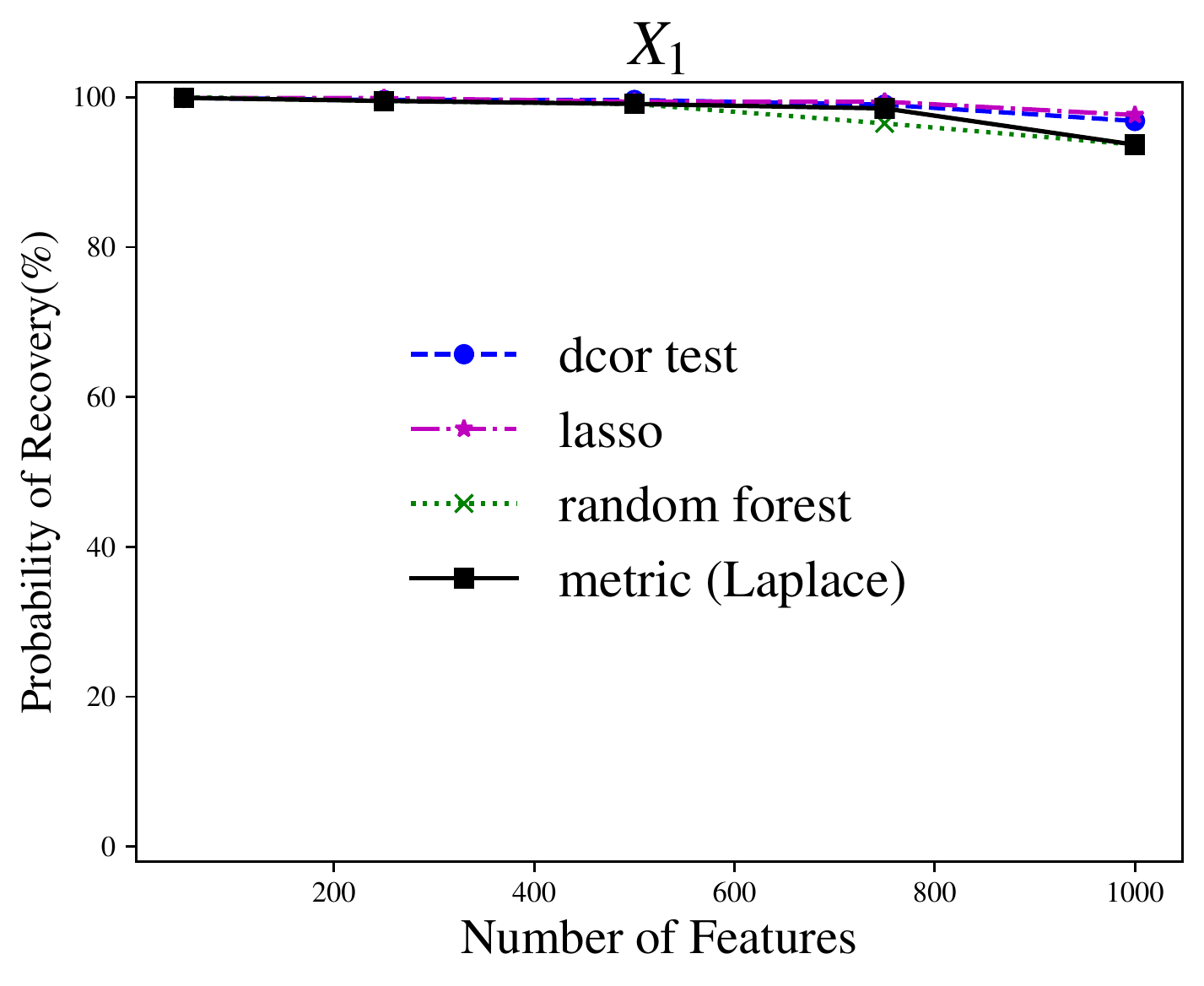}\includegraphics[width=.5\linewidth]{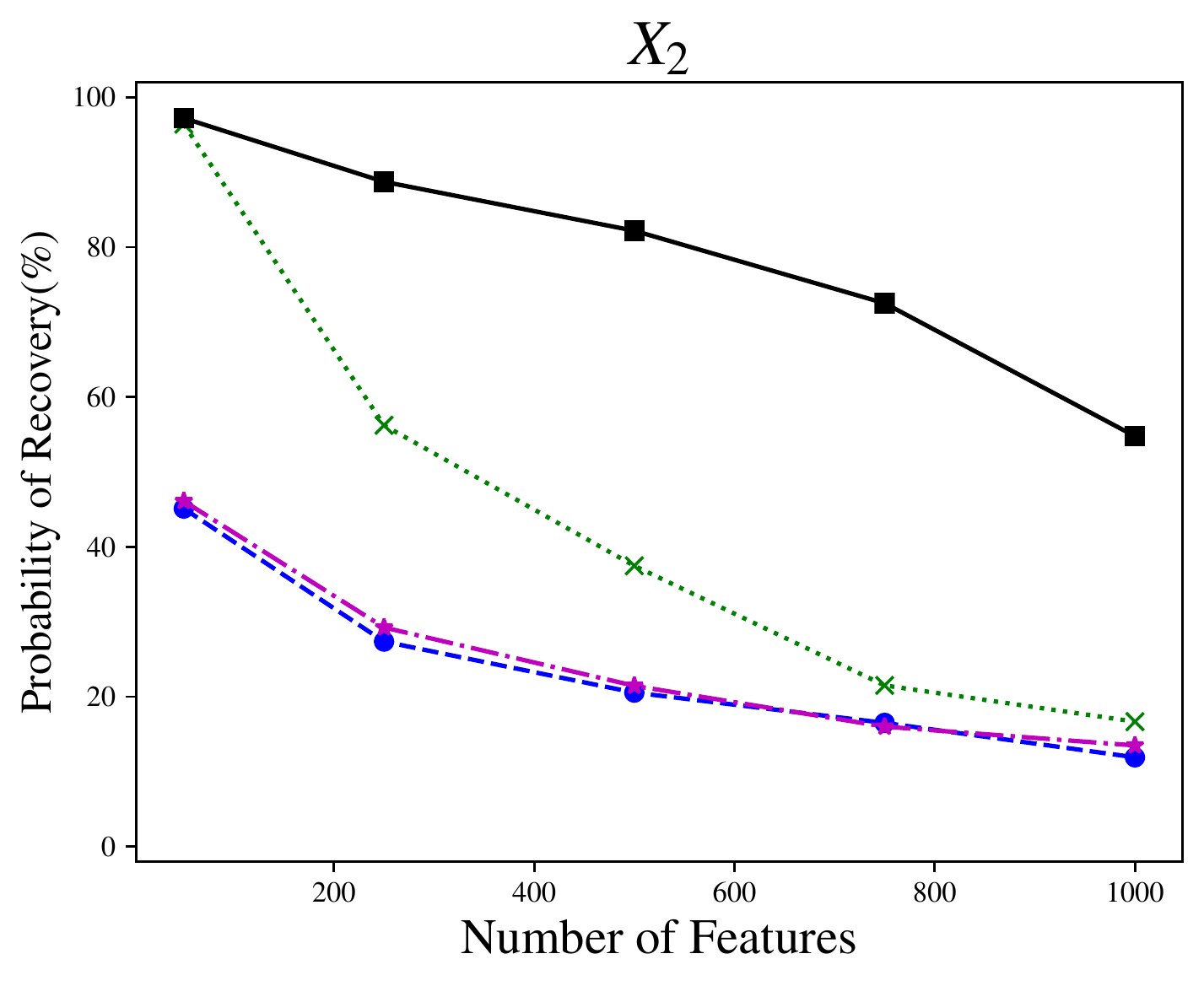}
\par\end{centering}
\begin{centering}
\includegraphics[width=.5\linewidth]{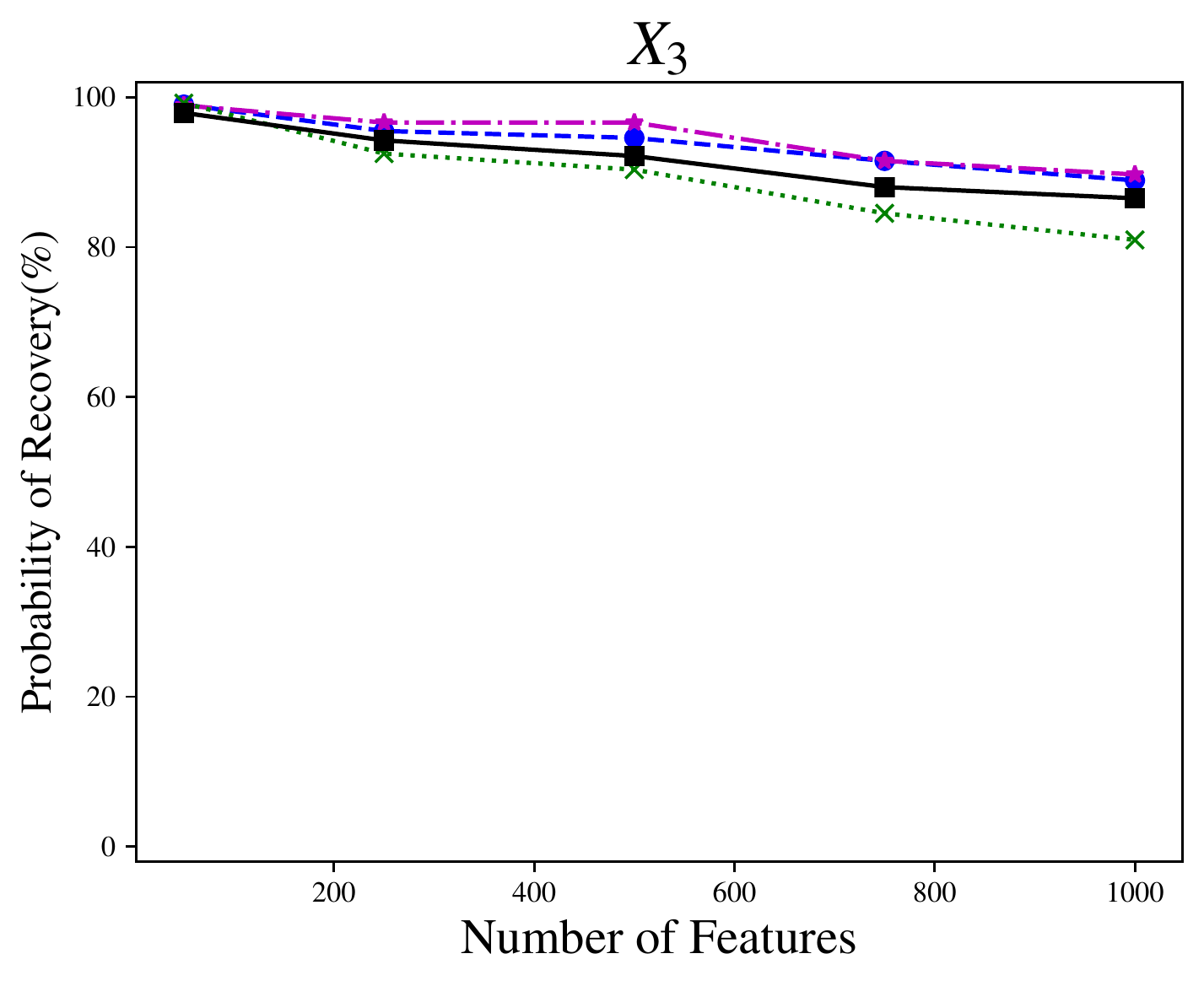}\includegraphics[width=.5\linewidth]{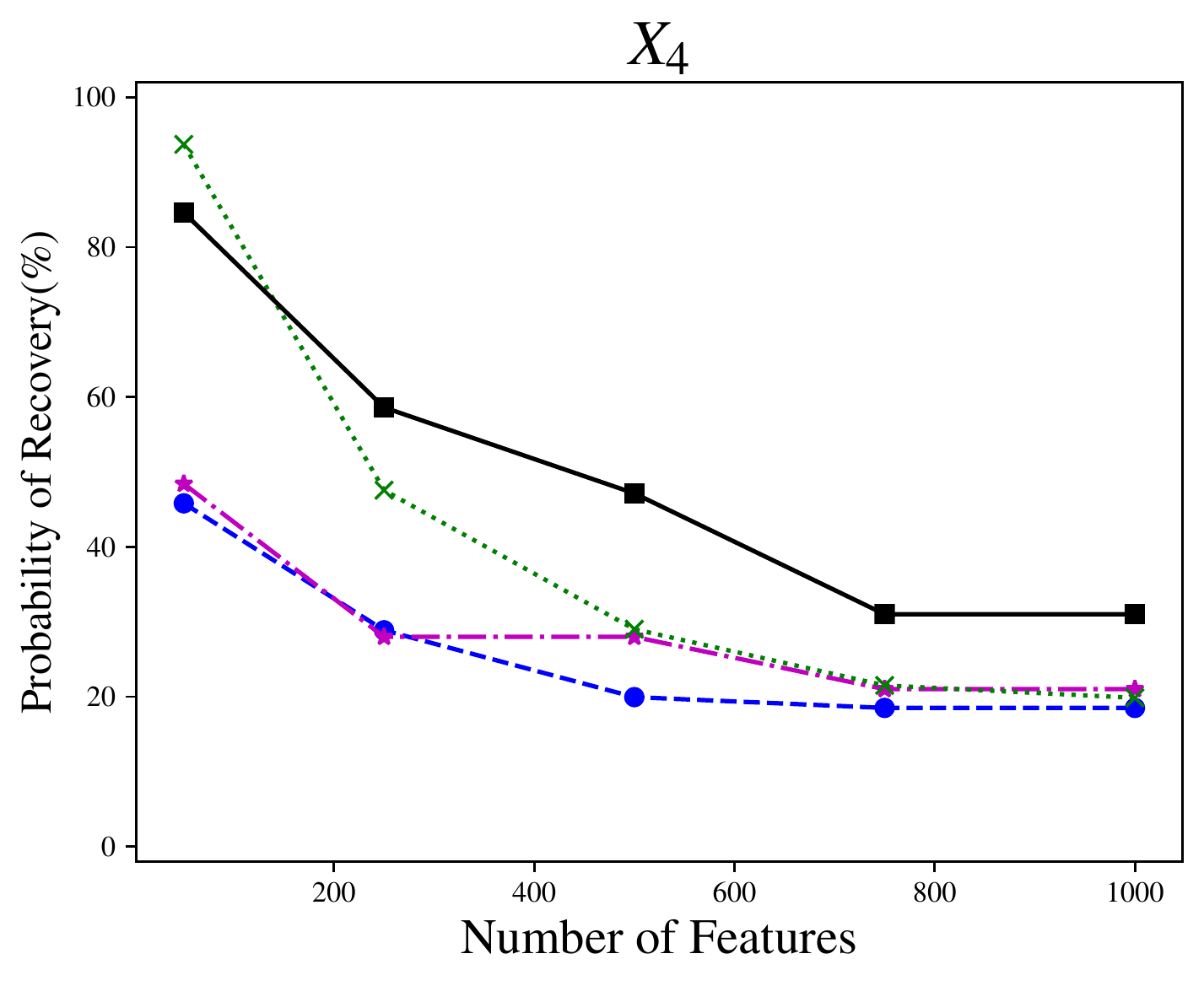}
\par\end{centering}
\caption{\small Probability of recovering variables involved in hierarchical
interactions in a QDA model (equation~\ref{eq:qda_model}). The sample size
$n=500$ is fixed while the number of null variables ranges over $\left\{ 50,250,500,750,1000\right\} $.}

\label{fig:qda}
\end{figure}

\begin{figure}[!tph]
\begin{centering}
\includegraphics[width=.5\linewidth]{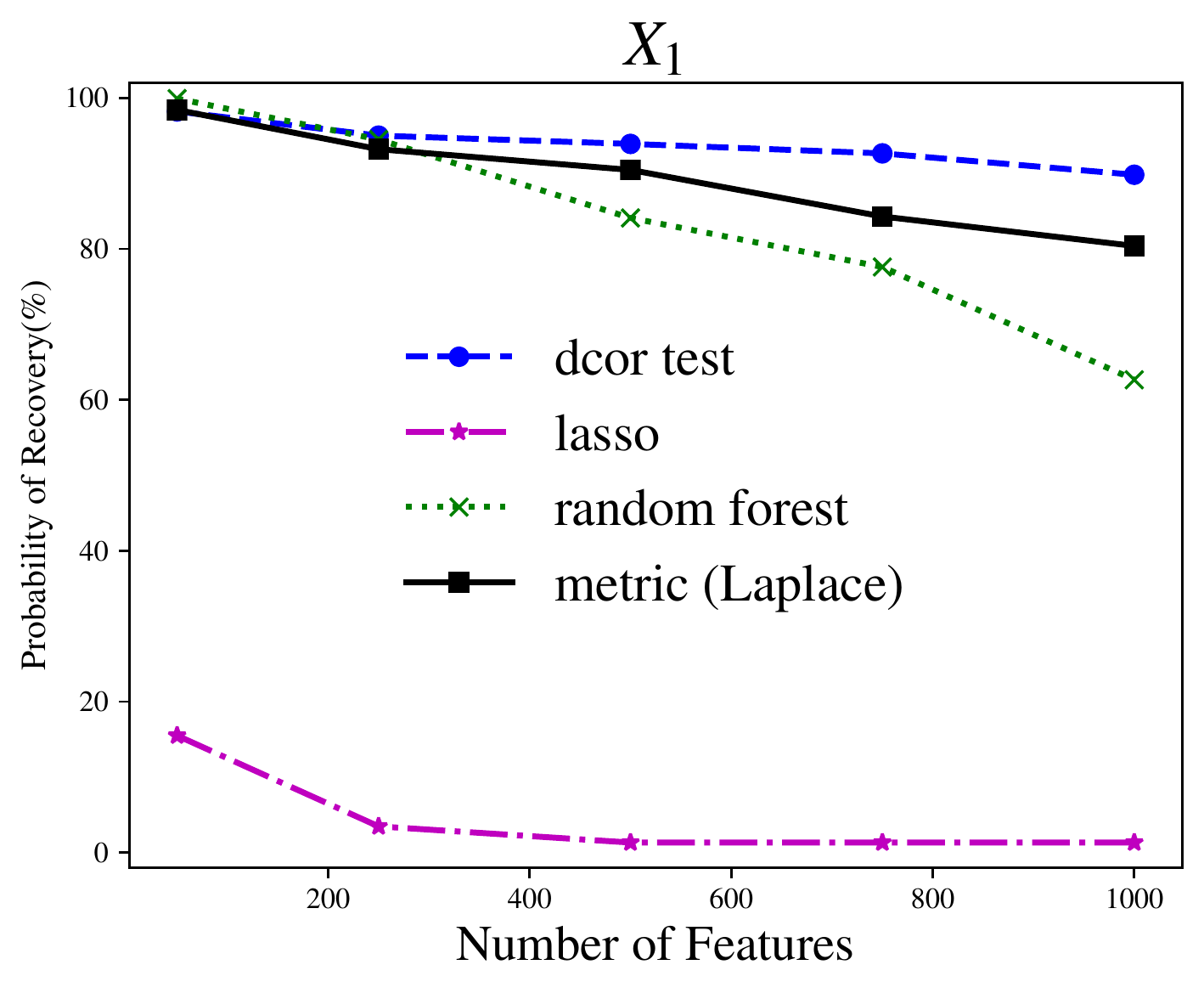}\includegraphics[width=.5\linewidth]{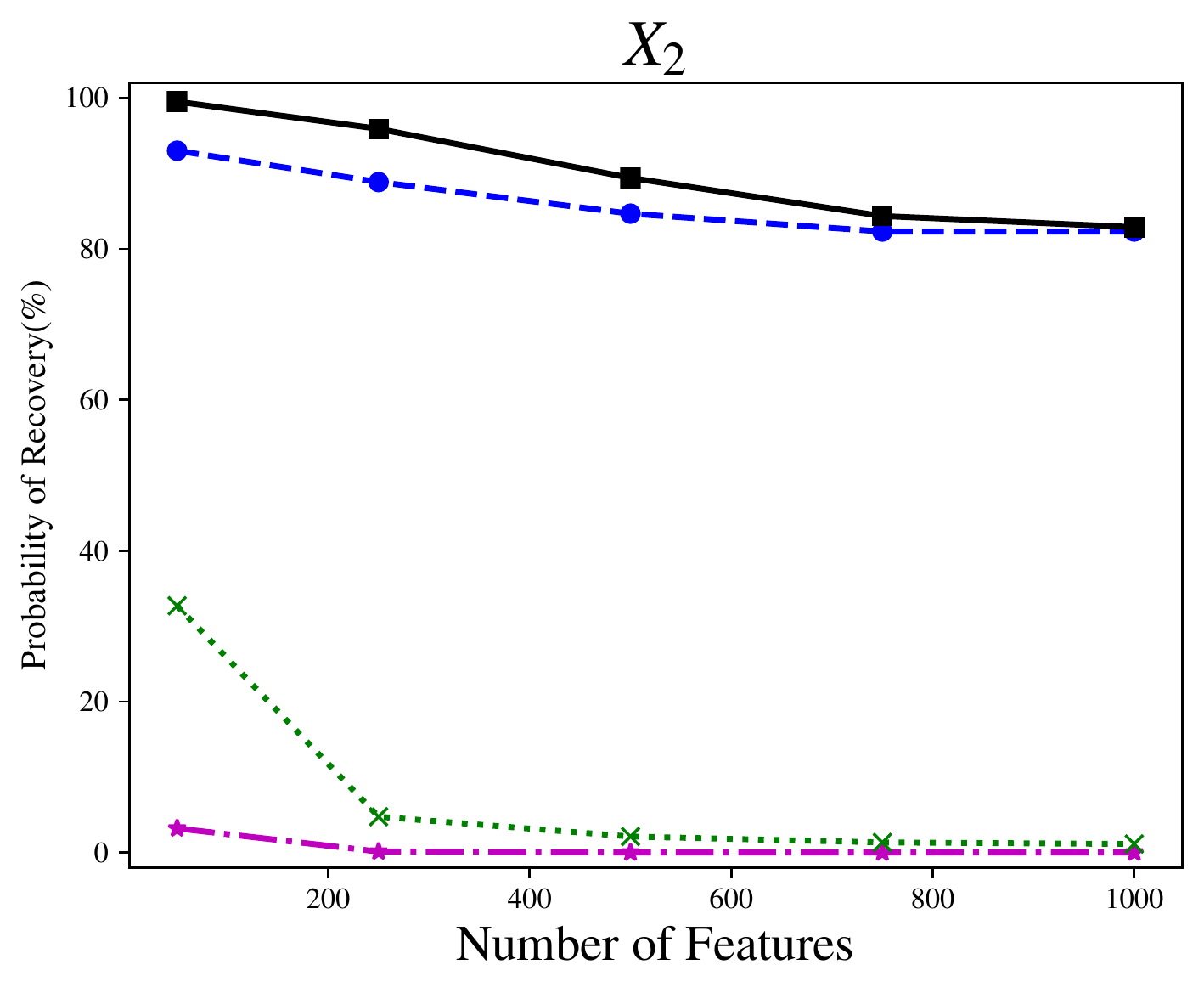}
\par\end{centering}
\begin{centering}
\includegraphics[width=.5\linewidth]{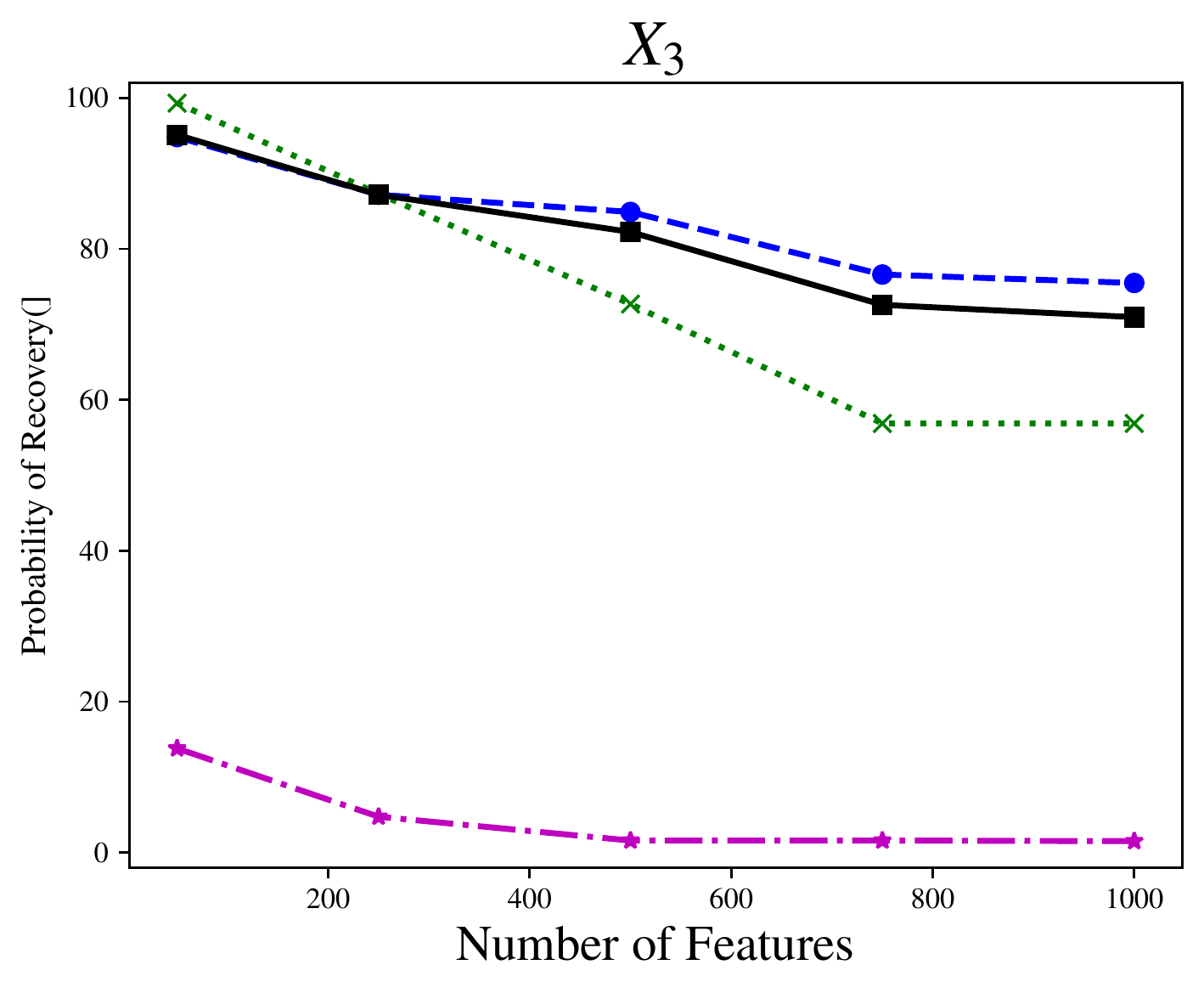}\includegraphics[width=.5\linewidth]{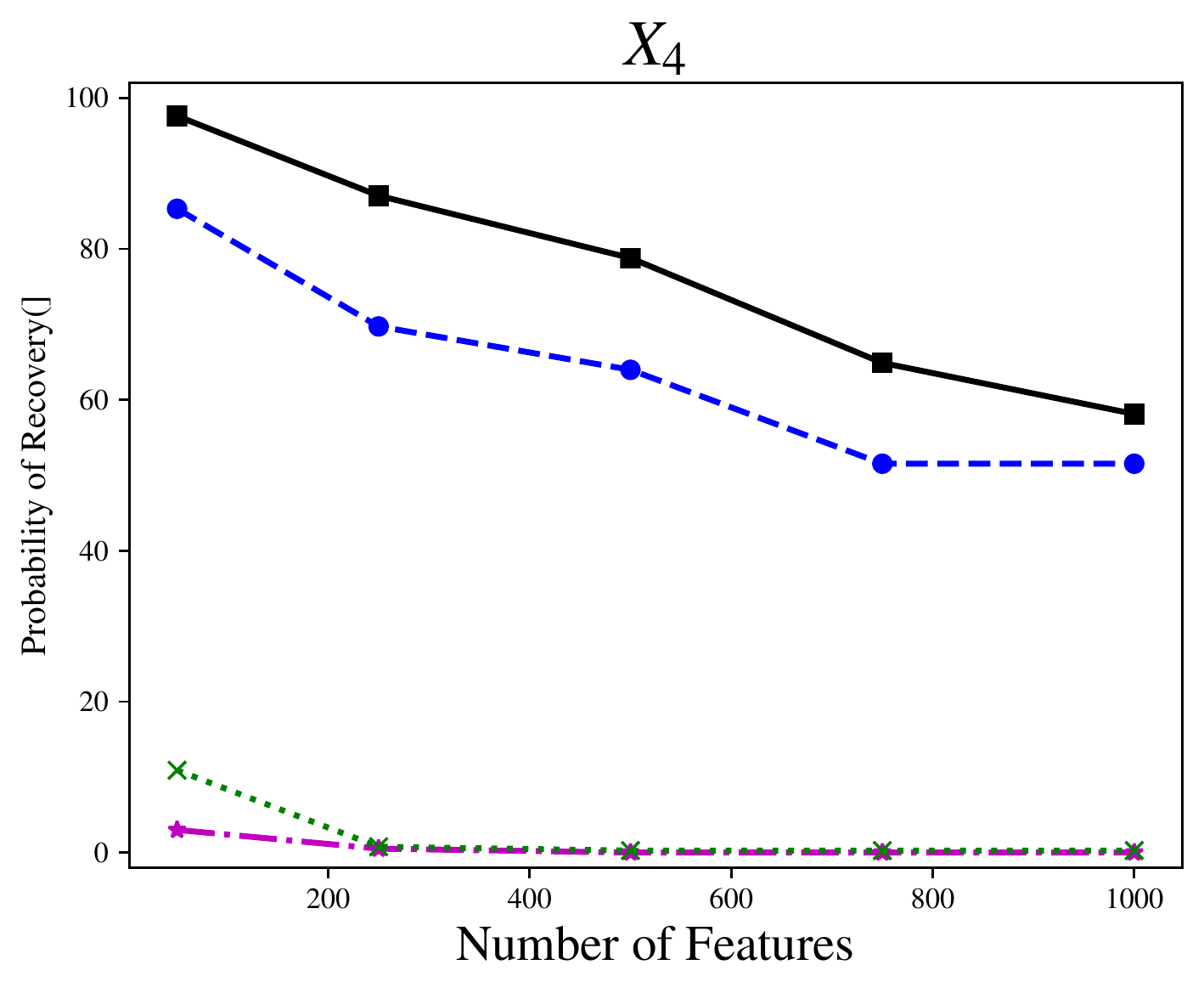}
\par\end{centering}
\caption{\small Probability of recovering variables in a logistic model with
ratio interactions (equation~\ref{eq:ratio}). The sample size $n=1500$
is fixed while the number of null variables ranges over $\left\{ 50,250,500,750,1000\right\} $.
The class balance is $\P(Y = +1) \approx 0.87$ and $\P(Y = -1) \approx 0.13$.}

\label{fig:ratio}
\end{figure}

We consider a logistic model where the signal is based on ratios of features:
\begin{equation}
\text{logit}~\P\left(Y=1|X\right)=\frac{\left|X_{2}\right|}{\left|X_{1}\right|}+0.8\cdot\frac{\left|X_{4}\right|}{\left|X_{3}\right|}.\label{eq:ratio}
\end{equation}
All features $X_j$ (signal and nosie) are independent $\normal\left(0,1\right)$. Here, $X_{1}$
and $X_{3}$ have strong main effects on $Y$ since they appear in
the denominator (hence are more influential) while $X_{2}$ and $X_{4}$
have weak main effects. $X_1$ interacts with $X_2$ and $X_3$ with $X_4$. Paralleling the story told by Figure \ref{fig:qda},
Figure \ref{fig:ratio} shows that metric screening is more effective than random forest in exploiting the existence of a hierarchical interaction to recover the variables with weak main effects $X_{2}$ and $X_{4}$.

\subsection{Pure Interaction: XOR}

\begin{figure}[!htb]
\label{fig:xor_2d}
\begin{centering}
\includegraphics[width=.58\linewidth]{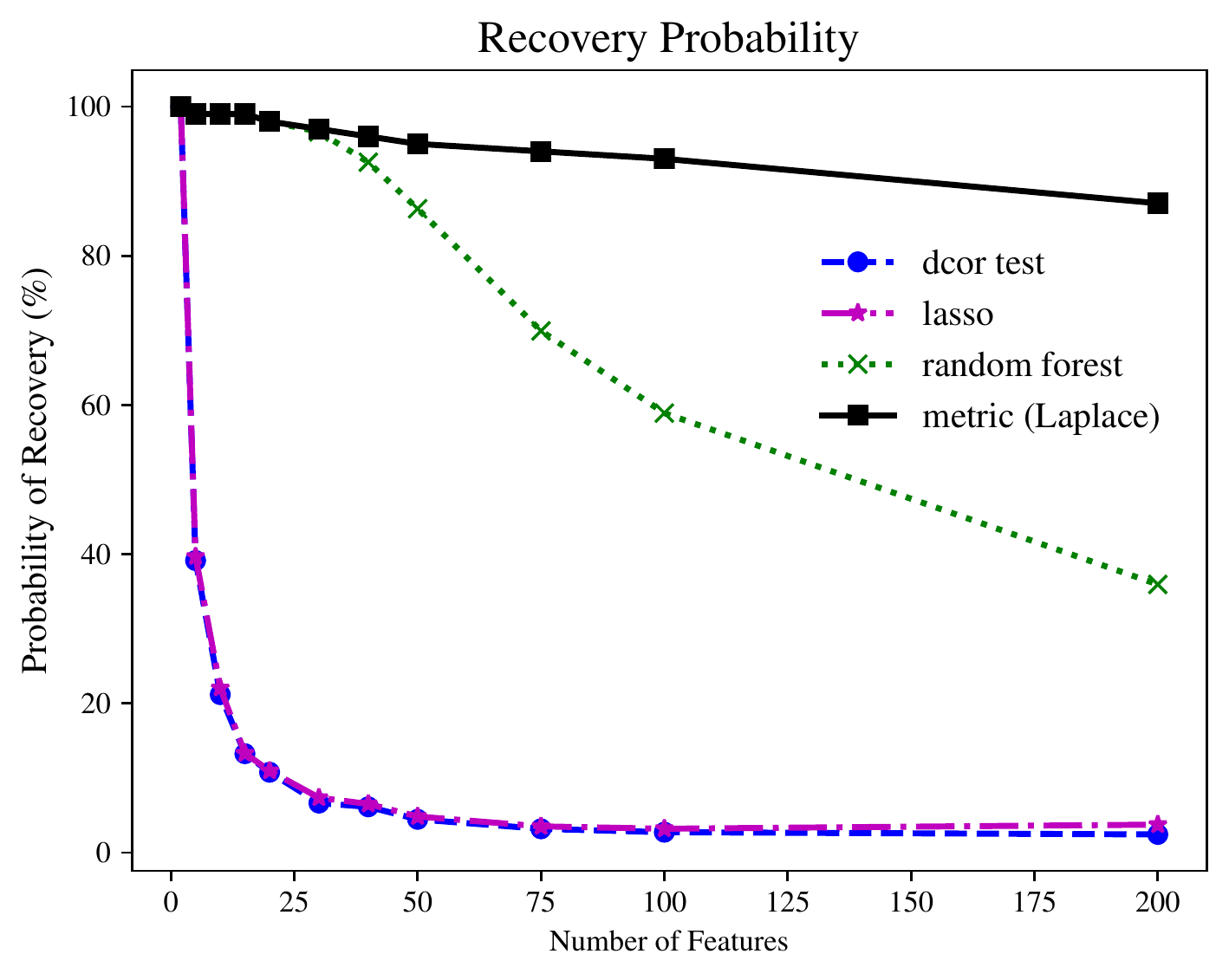}
\par\end{centering}
\caption{\small Impact of variable selection strategy on classification performance
of 2-layer fully connected neural network with 100 hidden units. The underlying model is an XOR signal, $Y=1_{X_{1}X_{2}>0}$,
where $X_{j}$ are independent $\normal(0, 1)$ for $1\le j\leq p$.
The number of features $p$ ranges from $2$ to $200$ (there are $p-2$
noise variables) and sample size is fixed at $n=1000$. Metric screening
allows neural networks to remain effective in higher dimensions.}
\end{figure}

We consider the classic XOR signal: $Y = \mathbb{I}_{X_1 \cdot X_2 > 0}$ where all variables $X_j$ (signal and noise) are i.i.d. 
$\normal(0,1)$. Both $X_1$ and $X_2$ are marginally independent of the response but together exhibit a pure interaction. 
Figure \ref{fig:xor_2d} plots the simulation results. Of the methods, only metric screening is built to handle pure interactions 
and as expected, it does the best.

\section{Conclusions}

\label{sec:conclusion}

Penalized linear models have been the workhorse of variable selection: they are
interpretable, cheap to fit, and work well in high dimensions. Their
great weakness is an inability to \emph{elegantly} model interactions: To capture order $s$ interactions
requires a model with $p^{s}$ terms! The class of metric learning objectives is a great nonparametric alternative to the linear model that allows detection of interactions at a linear computation cost.

\section*{Acknowledgements}
The authors thank Yuxin Chen, John Duchi, Michael Jordan, Xiao-li Meng, Weijie Su and Yichen Zhang for helpful
discussion and feedback. Robert Tibshirani started us on our journey to find a computationally efficient interaction detection 
algorithm and we could not have finished the journey without the encouragement and insights he provided throughout.

\section*{Appendix}
The appendix can be found at~\url{https://fengruan.github.io/pdf/paper-interactions-appendix.pdf}.

\bibliography{bib}

\bibliographystyle{abbrvnat}

\end{document}